\documentclass[11pt]{article}
\usepackage{times}
\usepackage{geometry}
\usepackage[utf8]{inputenc}
\usepackage{acronym,amsmath,amssymb,authblk,booktabs,caption,comment,enumitem,graphicx,lineno,mathtools,multicol,ragged2e,subcaption,upgreek,url,wrapfig,xspace}

\setenumerate{itemsep=0mm}
\usepackage[usenames]{xcolor} 
\definecolor{xlinkcolor}{cmyk}{1,1,0,0}

\usepackage[sort&compress,numbers]{natbib}

\usepackage{siunitx}
\usepackage[version=4]{mhchem}

\usepackage[
 colorlinks=true,    
 linkcolor=xlinkcolor,     
 citecolor=xlinkcolor,     
 filecolor=xlinkcolor,  
 urlcolor=xlinkcolor,      
 final=true
]{hyperref}

\newcommand{\affiliation}[1]{\textemdash\textit{#1}}

\newacro{NME}{nuclear matrix element}
\newcommand{\sm}{Standard Model\xspace}

\newcommand{\cross}{CROSS\xspace}
\newcommand{\cuore}{CUORE\xspace}
\newcommand{\cuoreo}{CUORE-0\xspace}
\newcommand{\cuoricino}{CUORICINO\xspace}
\newcommand{\cupid}{CUPID\xspace}
\newcommand{\cupido}{CUPID-0\xspace}
\newcommand{\cupidmo}{CUPID-Mo\xspace}
\newcommand{\cupidt}{CUPID-1T\xspace}

\newcommand{\qis}{QIS\xspace}

\newcommand{\dbd}{$2\mathrm{\nu\beta\beta}$\xspace}
\newcommand{\dbdgen}{double-beta decay\xspace} 
\newcommand{\nord}{normal ordering}

\newcommand{\ndbd}{$0\mathrm{\upnu\upbeta\upbeta}$\xspace}
\newcommand{\CEvNS}{CE$\upnu$NS\xspace}
\newcommand{\mbb}{$m_{\beta\beta}$\xspace}

\newcommand{\thalf}{$T_{1/2}^{0\nu}$\xspace}
\newcommand{\ckky}{counts$/($keV$\cdot$kg$\cdot$yr$)$}

\newcommand{\Mo}{\ce{^{100}Mo}\xspace}
\newcommand{\Te}{\ce{^{130}Te}\xspace}

\newcommand{\XeES}{$^{130}$Xe\xspace}
\newcommand{\lmo}{\ce{Li_2MoO_4}\xspace}
\newcommand{\moothree}{\ce{MoO_3}\xspace}
\newcommand{\enrlmo}{\ce{Li_{2}{$^{100}$}MoO_{4}}\xspace}

\newcommand{\teo}{\ce{TeO_{2}}\xspace}

\newcommand{\LiCo}{Li$_2$CO$_3$\xspace}
\newcommand{\Tc}{\ce{^{100}Tc}\xspace}
\newcommand{\Ru}{\ce{^{100}Ru}\xspace}

\newcommand{\al}{$\alpha$\xspace}
\newcommand{\be}{$\beta$\xspace}
\newcommand{\ga}{$\gamma$\xspace}

\DeclareSIUnit{\counts}{cts}
\DeclareSIUnit{\year}{yr}
\DeclareSIUnit{\dru}{\counts\per\kilo\electronvolt\per\kilogram\per\year}
\DeclarePairedDelimiterXPP\BigOSI[2]{\mathcal{O}}{(}{)}{}{\SI{#1}{#2}}

\newcommand{\RT}[1]{\textbf{\textcolor{red}{#1}}} 
\newcommand{\pingem}[1]{\textbf{\textcolor{orange}{#1}}} 

\newcommand{\qtd}[1]{\textcolor{purple}{#1}} 

\newcommand{\comm}[1]{\textcolor{black}{#1}} 

\newcommand{\commtoo}[1]{\textcolor{black}{#1}} 

\title{Toward \cupidt{}}
\author{CUPID Collaboration}
\date{\today}

\begin{document}
\maketitle

\begin{abstract}

Current experiments to search for broken lepton-number symmetry through the observation of neutrinoless double-beta decay (\ndbd{}) provide the most stringent limits on the Majorana nature of neutrinos and the effective Majorana neutrino mass (\mbb{}). The next-generation experiments will focus on the sensitivity to the \ndbd\ half-life of $\mathcal{O}(10^{27}$--$10^{28}$~years$)$ and $m_{\beta\beta}\lesssim15$~meV, which would provide complete coverage of the so-called Inverted Ordering region of the neutrino mass parameter space. 
By taking advantage of recent technological breakthroughs, new, future calorimetric experiments at the 1-ton scale can increase the sensitivity by at least another order of magnitude, exploring the large fraction of the parameter space that corresponds to the Normal neutrino mass ordering. In case of a discovery, such experiments could provide important insights toward a new understanding of the mechanism of \ndbd.

\comm{We present here a series of projects underway that will provide advancements in background reduction, cryogenic readout, and physics searches beyond \ndbd, all moving toward the next-to-next generation CUPID-1T detector.} 


\end{abstract}

\tableofcontents

\clearpage

\section{Introduction to \cupidt{}}

Cryogenic calorimeters (bolometers) play a unique role in the search for rare events and new processes, in which sharp resolution and low backgrounds enable sensitivity to weak scale interactions that can only be seen 
\commtoo{with very sharp energy resolution. 
Such detectors are ideal in the search for neutrinoless double-beta decay.  Detectors capable of reaching very low energy thresholds have also played a major role in the search for dark matter, where the complementarity of data from these experiments to data obtained from collider, accelerator, and satellite experiments has provided important constraints.  In recent years, breakthroughs in high-purity crystal production have enabled the scalability of large crystal experiments, enabling higher exposures for greater search sensitivities.
}

\paragraph{Neutrinoless Double Beta Decay (\ndbd)} Nuclear beta decay is a fairly common process involving the decay of neutrons to protons (or protons to neutrons) within the nucleus, resulting in the creation and ejection of an electron (or positron), and an antineutrino (or neutrino) and often accompanied by one or more gammas.  Similar processes include electron capture, where an electron (generally from one of the inner shells) is captured by the nucleus and results in a similar transmutation and accompanying release of energy \cite{DellOro:2016tmg}.
In cases where beta decay is suppressed by energy conservation or selection rules, less common, second-order processes may be observed.  In double-beta decay, two neutrons decay into two protons (or vice versa), resulting in a daughter nucleus of atomic number $Z+2$ (or ($Z-2$)) and a pair each of electrons (or positrons) and antineutrinos (or neutrinos).
This process has been observed in several isotopes, with half-lives of order 10$^{18}$--10$^{24}$ \si{year}.  Double-beta decays typically occur in near-stable isotopes with even numbers of both neutrons and protons, and occur slightly more frequently in isotopes with a greater number of neutrons, due to pairing and symmetry effects on the binding energy.  \cite{Wong:1990}.  

One of the most interesting aspects of measurements of double-beta decay is its implications for the origin and  nature of the neutrino mass.  In contrast to the \sm of particle physics, where neutrinos are nominally massless 
\cite{Elliott:2004, Dolinski:2019}, measurements of oscillations between the three known neutrino flavors over the last several decades have demonstrated the existence of the neutrino mass and characterized the relative splitting between mass eigenstates.     
One mechanism for generating the mass and also explaining the apparent lightness of neutrinos relative to other fermions is through the so-called see-saw mechanism \cite{PhysRevLett.44.912}.  In this scenario, the light neutrinos may have a Majorana mass component, leading to the possibility that neutrinos can act as their own antiparticles \cite{DellOro:2016tmg}.  As a result, isotopes which undergo \dbd may also be suceptable to a much rarer process known as neutrinoless \dbd (\ndbd), in which the interaction of the two neutrinos in the \dbd process results in their disappearance and the production of two betas only, with a summed energy at the Q-value of the interaction \cite{PhysRev.48.512}.
Such a scenario would have several important ramifications for neutrino physics and for the \sm of particle physics.  One striking consequence for particle physics is that the process would be a violation of lepton number symmetry, a beyond-\sm effect, and could play a vital role in the mechanism for the observed matter-antimatter asymmetry \cite{DellOro:2016tmg}.

\begin{figure}[hbpt]
    \centering
    \includegraphics[width=0.8\textwidth]{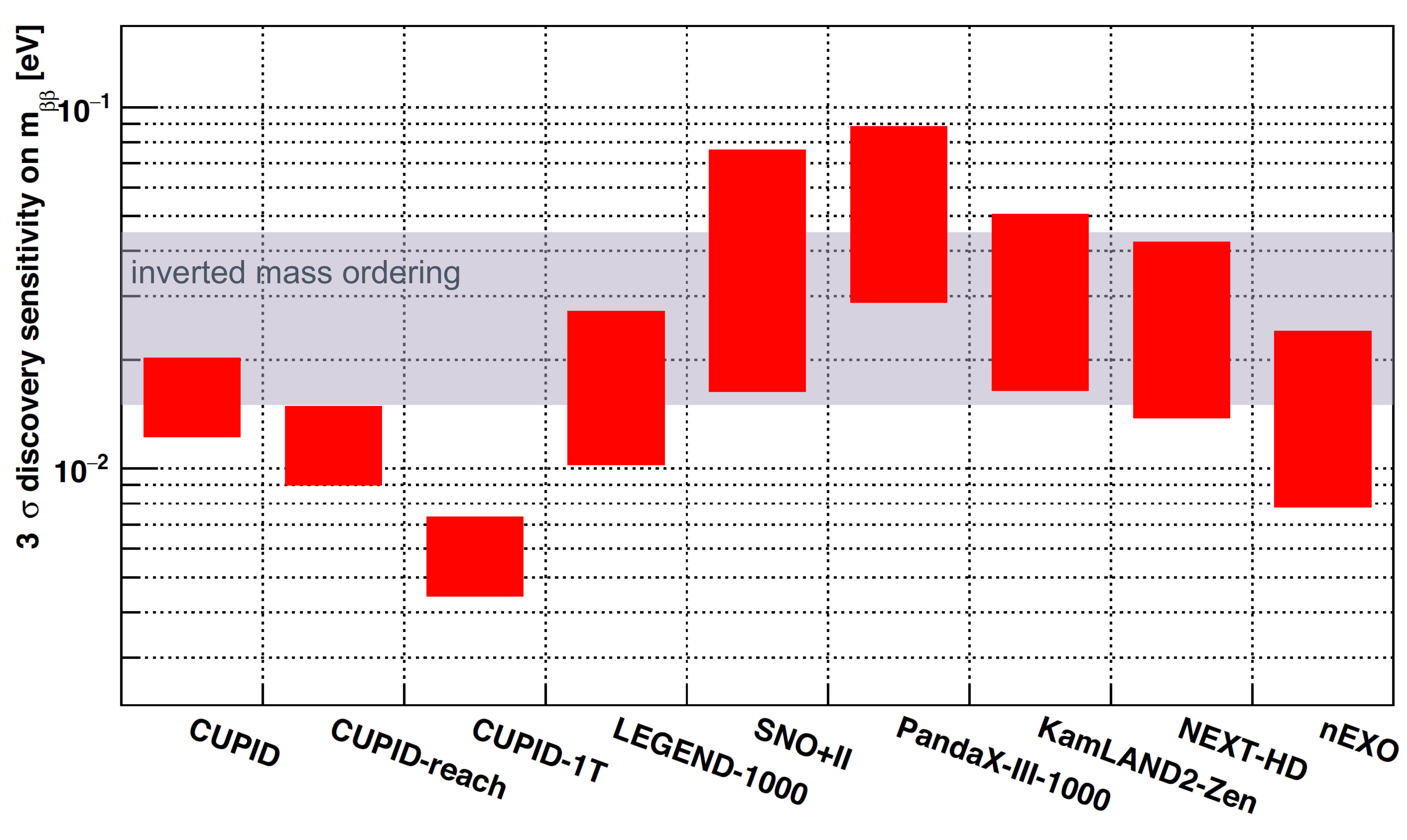}
    \caption{
    Discovery sensitivity for a selected set of next-generation ton-scale experiments.  The grey shaded region corresponds to the parameter region allowed in the Inverted Ordering of the neutrino mass.  The red error bars show the $m_{\beta\beta}$ values such that an experiment can make at least a $3\sigma$ discovery, within the range of the nuclear matrix elements for a given isotope. (\emph{Preliminary.})
    }
    \label{fig:2021sensitivitycomparison}
\end{figure}

\begin{figure}[hbpt]
    \centering
    \includegraphics[width=5in]{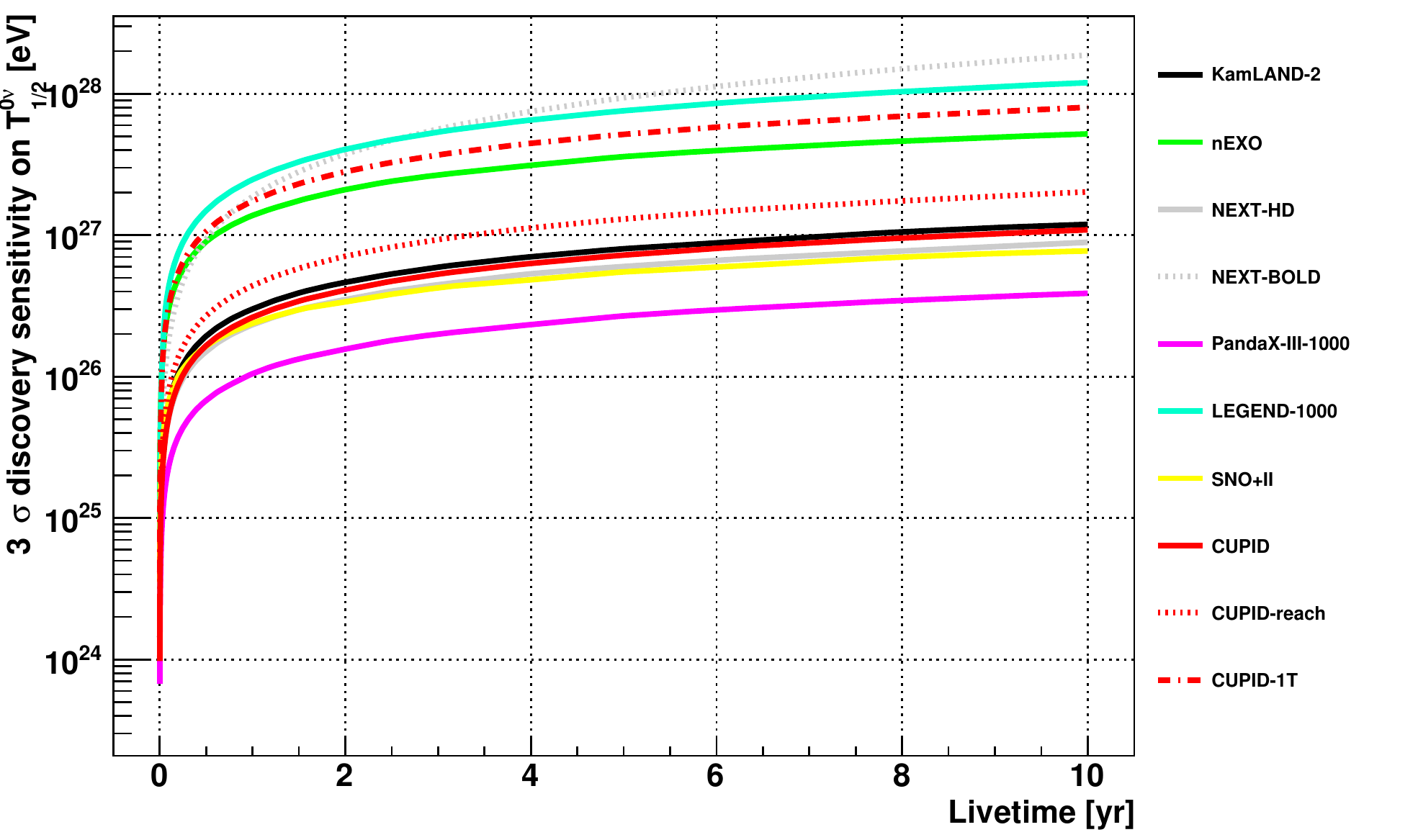}
    \caption{The red dot-dash curve corresponds to the CUPID-1T 3$\sigma$ discovery sensitivity across various ongoing and proposed experiments. CUPID-1T is expected to reach almost 10$^{28}$ yr sensitivity.  \emph{(Figure from the CUPID pre-CDR \cite{cupid2019}.)}}
    \label{fig:2019exclsenspCDR}
\end{figure}

Multiple experimental efforts have been proposed or planned worldwide to search for \ndbd. Fig.~\ref{fig:2021sensitivitycomparison} shows the $3\sigma$ discovery sensitivities in terms of \mbb\ for several proposed next-generation \ndbd decay experiments, including the CUPID program (with baseline and reach sensitivities), assuming a livetime of 10~yr for each experiment.  The \thalf\ sensitivity of each experiment (as shown in Fig. \ref{fig:2019exclsenspCDR}) is converted into a sensitivity on \mbb\ that depends on the \ac{NME} for the corresponding isotope. The uncertainty of each \ac{NME} is represented by the vertical extent of the bars in Fig.~\ref{fig:2021sensitivitycomparison}. For each isotope, we used the phase space factors from Ref.~\cite{Kotila:2012zza}, and all \acp{NME} available in literature~\cite{Barea:2015kwa,Simkovic:2013qiy,Hyvarinen:2015bda,Neacsu:2014bia,Menendez:2008jp,Rath:2013fma,Rodriguez:2010mn,Mustonen:2013zu,Meroni:2012qf,Vaquero:2014dna,Yao:2014uta,Horoi:2015tkc,Senkov:2014wtz}. The band for \Mo\ is narrower than for the other isotopes due to the lack of \acp{NME} computed with the interacting shell model.  A dedicated calculation has been requested by the \cupid\ Collaboration and is ongoing~\cite{Menendez}, while a recent work \cite{PhysRevC.105.034312} presents the shell model calculation for \ce{^{100}Mo} for the first time.  Figure~\ref{fig:2021sensitivitycomparison} shows that even with a modest amount of enriched isotope, \cupid\ is able to cover most of the region allowed in the inverted mass regime, even for the current largest \ac{NME} values. CUPID-1T extends the reach into the normal hierarchy, reaching almost 10$^{28}$ year sensitivity after a livetime of 10~yr (Fig. \ref{fig:2019exclsenspCDR}).

\commtoo{
\paragraph{From CUORE, to CUPID, to CUPID-1T}The \cuore{} line of \ndbd{} searches \cite{cuore1998, cuore:2022} uses large arrays of crystal calorimeters operated at temperatures of order \SI{10}{\milli\kelvin} to search for the neutrinoless double-beta decay of several key isotopes.  The \cuoricino{}, \cuoreo{}, and \cuore{} experiments were optimized to search for the decay of \ce{^{130}Te}. \cuore, the first ton-scale bolometric experiment, has demonstrated the capability of the modern cryogenic systems to operate the large detectors with high efficiency, excellent energy resolution, and low backgrounds. Moreover, the recent demonstrator experiments \cupido{} and \cupidmo{} placed world-leading limits on the decay of \ce{^{82}Se} and \cupidmo{}, respectively, while demonstrating the effectiveness of dual heat-light readout technique for particle identification and background suppression. The upcoming next-generation \cupid{} \cite{cupid2019} experiment, which plans to deploy about 300~kg of \ce{^{100}Mo} in the CUORE cryostat, builds on these capabilities.  
}

A future neutrinoless double-beta decay experiment with \SI{1000}{\kilo\gram} of the isotope of interest could potentially reach half-life sensitivities of order $10^{28}$ \si{\year} at the $3\sigma$ level \cite{Ejiri:2021}.  This corresponds to an effective Majorana neutrino mass (\mbb{}) in the range \SIrange[range-units=single,range-phrase=--]{4}{7}{\milli\electronvolt}, and discovery potential within the allowed region of the \nord{} of the neutrino masses. 
Current concepts toward development of such an experiment include \cupidt{}, a potentially multi-site, highly-segmented calorimeter with $\approx$\SI{1}{\tonne} of isotopic mass. 

With \SI{1}{\tonne} of isotope, and conservatively achievable improvements in sensor and detector technology, \cupidt{} could probe well into the region of \nord{} within the next ten years.  In the case of a discovery, this detector could be used for the parallel exploration of multiple isotopes in the same experimental volume (depending on the required isotopic mass).  This would allow a careful study of systematic effects and provide important insight into the variation of nuclear processes across isotopes.  Furthermore, such a detector would enable improved optimization of searches for a rich array of beyond-\sm processes, ranging from lepton number and CP violation to the interactions of detector nuclei with Majorons, axions, and weakly interacting massive particles (WIMPs) --- one or all of which may contribute to dark matter in the universe. 

Progress toward the technologies required for \cupidt{} 
has the potential to inspire advancements across multiple subfields, ranging from the development of new instrumentation to constraints on nuclear theory.  In addition, reapplication of the techniques and technology developed for use in \cupid{} has the potential to impact a number of outside fields.  


\subsection{Requirements for \cupidt}\label{sec:CUPID1T_Requirements}

Reaching the sensitivity for coverage of the Normal Ordering requires a continued  emphasis on background reduction and the development of a robust readout system for large macrocalorimeter arrays.  \cupidt{} will require a background index of approximately \SI{5e-6}{\dru}\textemdash\ a challenging but achievable goal 20 times lower than the conservative goal for the upcoming \cupid{} experiment. 
\comm{The addition of light collection to scintillating cryogenic calorimeters allows discrimination between $\alpha$ and $\beta/\gamma$ events by using the heat/light signal ratio, which suppresses backgrounds from degraded-$\alpha$s. The first pilot experiment for CUPID, CUPID-0, tested this with 26 ZnSe crystals read out using bolometric light detectors, and recorded a 99\% reduction in $\alpha$ background \cite{Azzolini:2019tta}. CUPID-Mo, using \lmo, produced similar results with excellent $\alpha$ rejection \cite{armengaudNewLimitNeutrinoless2021}. Further reduction in this background is supported by ongoing R\&D, as discussed later in this work (e.g. sections  \ref{sec:ActiveGammaVeto}\& \ref{sec:Synergies_DEMETER_MUX})}. 

In addition, \cupidt{} will require the rapid and reliable readout of over 10,000 channels, including both phonon sensors and the accompanying advanced light detectors, while minimizing the wiring exiting the cryostat \comm{(see sections \ref{TES_MKID},\ref{sec:Multiplexing},\& \ref{sec:CMOS_ASIC})}.  Additional requirements include the ability to acquire a sufficient amount of isotope for the experiment, and the cryogenic expertise to operate large, stable cryogenic systems for long periods of uninterrupted livetime.  Ideally, the expertise gained from the current \cuore{} experiment and partners in dark matter, CMB experiments, and \qis{} can be applied to systems with volumes up to 4 times larger than that of the current \cuore{} experiment, and potentially replicated in multiple underground laboratories across the world. 

\begin{table}[t!]
  \caption{Parameters of the \cupid\ detector in the conservative baseline scenario, in the optimistic background scenario requiring the use of new, but existing, technologies \comm{(CUPID-reach)}, and for CUPID-1T. 
  (\emph{Preliminary}.)}
  \label{tab:phases}
  \centering
  \begin{tabular}{lrrr}
    \toprule
    Parameter & CUPID baseline & CUPID-reach & CUPID-1T \\
    \midrule
    Crystal & \enrlmo & \enrlmo & \enrlmo \\
    Detector mass (kg) & 450 & 450 & 1871 \\ 
    $^{100}$Mo mass (kg) & 240 & 240 & 1000 \\ 
    Energy resolution FWHM (keV) & 5 & 5 & 5 \\ 
    Background index (\ckky) & $10^{-4}$ & $2\times10^{-5}$ & $5\times10^{-6}$ \\ 
    Containment efficiency & 78\% & 78\% & 78\% \\ 
    Selection efficiency  & 90\% & 90\% & 90\% \\ 
    Livetime (years) & 10 & 10 & 10 \\
    Half-life exclusion sensitivity (90\% C.L.) & $1.4\times10^{27}$ y & $2.2\times10^{27}$ y & $9.1\times10^{27}$ y \\
    Half-life discovery sensitivity ($3\sigma$) & $1\times10^{27}$ y & $2\times10^{27}$ y & $8\times10^{27}$ y\\
    \mbb\ exclusion sensitivity (90\% C.L.)     & 10--17~meV           & 8.4--14~meV          & 4.1--6.8~meV \\
    \mbb\ discovery sensitivity ($3\sigma$)     & 12--20~meV           & 9--15~meV          & 4.4--7.3~meV \\
    \bottomrule
  \end{tabular}
\end{table}

\subsection{Variations on a Theme:  Modifications to a 1-Ton Design for Optimal Sensitivity}

The movement of \ndbd{} experiments toward the 1-ton regime presents the opportunity to expand upon the classic segmented bolometer configuration in several ways that can increase both the sensitivity and robustness of searches. 

The basic detector design has typically included single or multitower arrays of nearly identical crystals of a single substrate, mounted in a single cryostat. 
All crystals are instrumented with identical sensors and read out on dedicated channels.  Insofar as a fiducial volume is defined, it has typically been implemented by identifying regions of the array more susceptible to noise and backgrounds, and optimizing analysis cuts on detectors in a core region.  These more sensitive detectors typically have identical design parameters to those located within the more susceptible regions, and are optimal for search primarily by location.

Cost, access, reproducibility, scalability, and other considerations have contributed to the classic basic detector design.  With the maturation of the experimental approach and advances in detector modeling, characterization, and analysis, we now have the tools to incorporate more complex design elements that can improve detector performance. These more complex elements may include the use of varying crystal sizes in interior regions of the detector, and the simultaneous deployment of multiple isotopes in a single experiment, \comm{which, in the case of discovery, would allow for a more robust constraint on the mechanism of \ndbd. For example, one proposal suggests that an array of 1300 light-emitting cryogenic calorimeters --- using \ce{Zn$^{82}$Se}, \ce{Li${}^{100}$MoO$_4$}, ${}^{116}$CdWO$_4$ and ${}^{130}$TeO$_2$ --- could reach sensitivities near or beyond 10$^{27}$ years in each isotope\cite{giulianiMultiisotopeNuBeta2018}.}

\comm{Elimination of the enrichment requirement could reduce the dependence of \ndbd~experiments on potentially rogue governments, avoid geopolitical risk, and better align the \ndbd~ program with U.S. economic interests.  At the scale of CUPID-1T, a bolometric experiment with natural TeO$_2$ (and advanced topological reconstruction, discussed in section \ref{sec:Synergies_DEMETER_MUX}) would reach the sensitivity goals outlined by the 2015 Long Range Plan\cite{LongRangePlanReachingHorizon20152015a}, without the need for isotopic enrichment. This is a unique feature of this technology among currently proposed \ndbd~ experiments.}

\subsection{Multiple Visions for the Expansion to \cupidt{}}
Beyond the efforts at background reduction and additional cryogenic technology that we detail later in this work, the realization of \cupidt{} could be achieved by building a larger cryostat to house the increased volume of crystals. This has the advantage of the outer crystals acting as shielding for the inner-most crystals; the outer "veto" layers of the crystal array could be made out of unenriched material or larger crystals to reduce costs. The design of such a cryostat 
would be a natural extension of the \cuore{} expertise. 

Alternatively, \cupidt{} could leverage international enthusiasm for bolometers to stage the experiment in multiple cryostats around the world or multiple cryostats at the same location. This topology mirrors what is currently being implemented for quantum computers based on superconducting qubits. We note that there are $\sim$10\,kg-scale demonstrator experiments like CUPID-Mo (Modane, France)~\cite{Armengaud:2019loe}, CROSS (Canfranc, Spain)~\cite{Zolotarova:2020suz}, and AMoRE (Yangyang, Korea)~\cite{Lee:2020rjh}, and single crystal R\&D is proceeding in the U.S., France, China, and Japan.  Such a diffuse staging of the experiment would naturally lead itself to a multi-target observatory and the possibility of multiple \ndbd{} isotopes and additional physics topics (see Section \ref{sec:physics}).


\section{Physics Beyond Neutrinoless Double-Beta Decay} \label{sec:physics}
        The development of a ton-scale, multi-site, calorimetric detector would provide an opportunity to perform a wide array of searches for physics beyond the Standard Model.  In fact, the discovery of \ndbd{} would be an immediate indicator of broken lepton number symmetry.
        In addition to shedding light on the mechanism of \ndbd{}, a ton-scale cryogenic calorimeter would be sensitive to searches for low-mass dark matter candidates, the neutrino magnetic moment (using external sources), solar axions, {\color{blue} 
         } symmetry (Lorentz, CPT) violations, Majorons, lightly-ionizing and fractionally charged particles, and could serve as an observatory for the study of coherent neutrino scattering and cosmic-ray muons. 
        
        
        
          
The very low background rate expected in \cupid promises competitive sensitivities to such processes.  Moreover, the short decay half-life of \ce{^{100}Mo} $2\nu\beta\beta$ and the large source mass will yield over $10^8$ $2\nu\beta\beta$ events in 1 year of data collection.  This will provide a remarkable opportunity to reach unprecedented sensitivity to deformations of the continuous two-neutrino spectrum shape.

\subsection{Decays to Excited Nuclear States}


In the basic case of a ``simple'' two-neutrino double-beta decay (\dbd{}), the best experimental sensitivity is generally for the transition of the daughter nucleus to the ground state. However, \dbdgen{} (in both 2$\nu$ and 0$\nu$ modes) may also occur to an excited state of the daughter nucleus. 
In the case of \ndbd, these transitions can disclose the exotic mechanisms (e.g. right-handed currents) that mediate the decay~\cite{TOMODA2000245,SIMKOVIC2002201}, while for \dbd\ they can provide unique insight to the details of the mechanisms responsible for the nuclear transition~\cite{AUNOLA1996133,SUHONEN1998124}.
From an experimental point of view, most of the interest is motivated by the fact that in a close-packed array like CUPID, the strong signature  provided by the simultaneous detection of one or two gammas can lead to an almost background-free search. 
In this respect, the transitions to 0$^+$ states are favored while states with larger spin (e.g. 2$^+$) are generally suppressed by angular momentum conservation.
The strategy adopted to study such decays with an array of cryogenic calorimeters exploits multiple coincidence patterns to select topological configurations characterized by a lower background contribution~\cite{CUORE02012503,Azzolini:2018oph,Adams:2018yrj,Fantini:2020tkm}. A number of 0$^+$ and 2$^+$ excited states of $^{100}$Mo are accessible to CUPID with unprecedented sensitivity~\cite{BARABASH20078}.

\subsection{The \dbd{} spectrum and probes of symmetry violation} \label{sec:2nuShapeStudies}

In the past few years, attention has been drawn to the fact that a precision study of \dbd{} (accurate measurement of the half-life, accurate determination of the total spectrum shape, single electron spectrum, and angular distribution) will allow the study of unusual properties of neutrinos and processes beyond the Standard Model. 
Thus, the presence of a bosonic component in a neutrino will lead to a change in the shape of the spectra (total and single electron) and a change in the half-life \cite{Barabash:2007gb,NEMO-3:2019gwo}. Participation in double beta decay of sterile neutrinos \cite{Bolton:2020ncv,Agostini:2020cpz}, violation of Lorentz invariance \cite{Nitescu:2020xlr,Nitescu:2020qnh}, the presence of right-handed currents \cite{Deppisch:2020mxv}, light exotic fermions \cite{Agostini:2020cpz} and strong neutrino self-interactions \cite{Deppisch:2020sqh} also leads to a change in the shape of the total $2\nu\beta\beta$ spectrum. In this case, it is very important to know very precisely the theoretical spectral shape of \dbd{}\comm{, which will require separate theoretical study.} 

\subsubsection{Symmetry (Lorentz, CPT) violations} \label{Symmetry Violations} 
\begin{figure}[htbp]
    \centering
    \includegraphics[width = 5.0in]{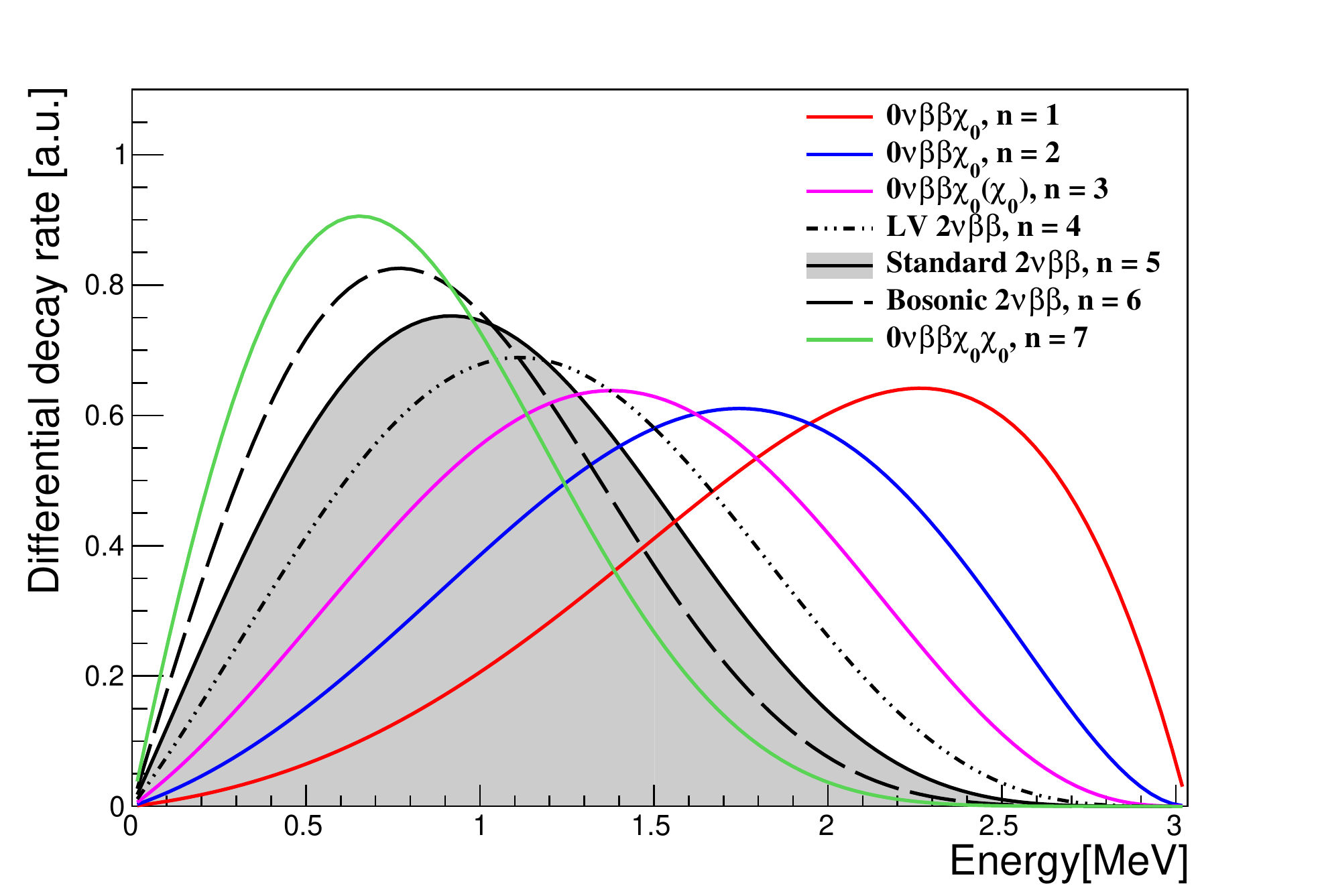}
    \caption{Differential decay rate for different decay modes: \dbd\ (the spectral index $n = 5$), single Majoron emission ($n$ = 1, 2, 3), double Majoron emission ($n$ = 3, 7), Lorentz Violation and in case of purely Bosonic neutrinos ($n$ = 6).}
    \label{fig:celi2nu}
\end{figure}

Lorentz invariance and CPT violations arising from the spontaneous breaking of the underlying space-time symmetry are interesting theoretical features that can be parameterized within the so-called Standard Model Extension (SME)~\cite{SME19976760,SME1998116002,SME2004105009}. 
Lorentz-Violating (LV) effects in the neutrino sector can appear both in the two-neutrino and in the neutrino-less decay mode~\cite{JORGE2014036002}. Indeed, a distortion of the two-electron summed energy is expected for \dbd\ due to an extra term in the phase space factor, while \ndbd\ could be directly induced by a Lorentz violating term. The signature is therefore very similar to the one expected for Majoron searches with a deformation of the upper part of the \dbd\ spectrum (fig.~\ref{fig:celi2nu}). The dominant contribution from \dbd of $^{100}$Mo, and the extremely low background expected above 2 MeV, make CUPID particularly sensitive to these effects. The parameter $\text{\textbf{\aa}}_{\text{of}}^{(3)}$ is related to the time-like component of the LV operator in the neutrino sector. The preliminary predicted upper limit of the Lorentz-violating term is $\text{\textbf{\aa}}_{\text{of}}^{(3)}<5.5 \times 10^{-8}$~GeV at 90\%~C.I. for an exposure equivalent to 1 year of \cupidt{} data taking. 
\subsubsection{Majorons}
\label{sec:Majorons}
Exotic \ndbd\ decays characterized by the emission of a massless Goldstone boson, called a Majoron, are predicted by some theoretical models~\cite{GELMINI1981411}. 
The precise measurements of the invisible Z width at LEP has greatly disfavoured the original Majoron triplet and pure doublet.
However, several new models involving massless or light bosons with a coupling to neutrinos have been developed~\cite{BURGESS19945925,BAMERT199525}. All of them predict different (continuous) spectral shapes for the sum energies of the emitted electrons, which extend from zero to the transition energy $Q_{\beta\beta}$. The differential decay rate can be approximated with the following formula: 
\begin{equation}
\frac{dN}{dT} \sim (Q_{\beta\beta}-T)^n,
\end{equation}
where $T$ is the electron summed kinetic energy and the spectral index, $n$, depends on the decay details. Single Majoron emissions are characterized by $n$ values between 1 and 3, while double Majoron decays can have either $n=3$ or $n=7$ (see Fig.~\ref{fig:celi2nu}). 
As for any process characterized by continuous spectra, the experimental sensitivity is mainly limited by the background contributions \comm{(in this case dominated by standard \dbd{} decay)} and the detector mass~\cite{ELLIOTT19871649,ARNOLD2006483,KZENMAJORON2012}. \comm{Fixing the background contributions of the CUPID preliminary background budget and scaling the activity of \Mo{} \dbd{}\cite{armatolSearchingNewPhysics2021},} \cupidt is expected to have excellent sensitivity to these processes, with values as reported in Table ~\ref{tab:majorons}\comm{; this is an order of magnitude or more improvement over the current limits in \Mo set by NEMO-3\cite{arnoldDetailedStudies1002019}}. Limits on the Majoron-neutrino coupling constant are set considering $1/T_{1/2} = |g_{ee}|^m G |M|^2$ where $G$ and $M$ are the phase space factor and the nuclear matrix element, respectively. Values of $G$ and $M$ come from theoretical calculations~\cite{Kotila:2021mtq,Kotila:2015ata}, while $m$ can be equal to 2 or 4 if we consider the emission of one or two Majorons. 
The Majoron is also seen as a possible dark matter candidate \cite{Rothstein:1992rh,Berezinsky:1993fm,Garcia-Cely:2017oco}. Recently this feature has attracted more and more attention, resulting in new models based on massive Majorons \cite{Blum:2018ljv,Cepedello:2018zvr,Brune:2018sab,Funcke:2019grs,Heeck:2019guh}. Some of these models can be tested by double-beta decay experiments \cite{Cepedello:2018zvr,Brune:2018sab}, in particular by \cupidt{}.\\
\begin{table}
\caption{List of decays with one ($\chi^0$) and two ($\chi^0\chi^0$) Majoron emission. Predicted limits on the half-lives and on Majoron-neutrino coupling $g_{ee}$ are at 90\% C.I., corresponding to an exposure of 1 year of data taking.}
\centering
\begin{tabular}{cccc}
\toprule
$n$ & mode & predicted lower limit on T$_{1/2}$ [yr] & $g_{ee}$ exclusion sensitivity\\
\midrule
1 & $\chi^0$ & 1.3 $\times$ 10$^{24}$ & (4.7 - 6.6) $\times$10$^{-6}$ \\
2 & $\chi^0$ & 2.0 $\times$ 10$^{23}$ & - \\
3 & $\chi^0$ & 6.7 $\times$ 10$^{22}$ & (0.8 - 1.4) \\
3 & $\chi^0\chi^0$ & 2.7 $\times$ 10$^{22}$ & (4.0 - 5.9) $\times$10$^{-3}$\\
7 & $\chi^0\chi^0$ & 1.2 $\times$ 10$^{22}$ & (3.8 - 7.1) $\times$10$^{-1}$\\
\bottomrule
\end{tabular}
\label{tab:majorons}
\end{table}


\subsection{Other Spectral studies}\label{sec:otherspec}
In addition to CPT symmetry, there are other conservation laws whose violation produces  experimental signatures recognizable from CUPID.

\paragraph{Pauli Principle violation} 
        The Pauli exclusion principle (PEP) is one of the basic principles of physics upon which modern atomic and nuclear physics are built. 
        Despite its well known success, the exact validity of PEP is still an open question and experimental verification is therefore extremely important~\cite{GREENBERG198983}. 
        Indeed, a number of experimental investigations have been carried out both in the nuclear and atomic sector~\cite{BXINO2010034317,VIP2018319,BELLI1999236}. 
        In all the cases, the signature is a  transition between already-occupied (atomic or nuclear) levels, which is clearly prohibited by PEP. Most of the low activity experiments exploit large masses and/or low background rates to search for the emission of specific electromagnetic or nuclear radiation from atoms or nuclei. In contrast, dedicated PEP-violation searches aim at improving the sensitivity by filling already complete atomic levels with fresh electrons and measuring the corresponding X-ray transitions. Unfortunately, a model linking the two experimental observations is still missing and a comparison of the sensitivities is impossible. 
        CUPID belongs to the first category and will exploit the excellent energy resolution and the very low background to search for the emission of X-rays, $\gamma$s, or nucleons from the detector atoms and/or nuclei. 
        PEP-prohibited nuclear decays are also possible, although a lower experimental sensitivity is anticipated for CUPID. 

\paragraph{Tri-nucleon decays} \label{sec:OtherSpectra_TriNucleonDecays}
    Baryon number (B) conservation is an empirical symmetry of the SM.
    Its violation is predicted by a number of SM extensions. Furthermore, it is expected that quantum gravity theories violate B and that theories with extra dimensions permit nucleon decay via interactions with dark matter~\cite{BABU20135285}. 
    In particular, some SM extensions, which allow for small neutrino masses, anticipate $\Delta$B=3 transitions in which three baryons can simultaneously disappear from the nucleus, frequently leaving an unstable isotope~\cite{BABU200332}. The coincidence between the tri-nucleon decay and the radioactive decay of the daughter nuclei is then a robust signature which can help mitigate the backgrounds. The dominant $\Delta$B=3 decay modes are $ppp \to e^+ \pi^+ \pi^+$, $ppn \to e^+ \pi^+$, $pnn \to e^+ \pi^0$, and $nnn \to \bar{\nu} \pi^0$. The decay-mode-specific signatures (charged fragments) include an initial saturated event followed by one or more radioactive decays. The invisible decay-mode signatures are composed of two successive decays and hence have two energy constraints and one time constraint.

    \paragraph{Electric charge conservation.} 
        The decay of an atomic electron is probably the most sensitive test of electric charge conservation.
        Charge non-conservation (CNC) can be obtained by including additional interactions of leptons and photons which lead to the decay of the electron: $e \to \gamma \nu$ or $e \to \nu_e\nu_X\bar{\nu}_X$. 
        These modes conserve all known quantities apart from electric charge.
        In addition, CNC can also involve interactions with nucleons. Discussions of CNC in the context of gauge theories can be found in a number of BSM gauge models~\cite{VOLOSHIN1978145,OKUN1978597,IGNATIEV1979315}.
        While the signature of the neutrino mode for CNC is quite poor, the coincidence between the gamma and the X-rays emitted via atomic de-excitations can give rise to interesting topological configurations that can help to lower the background contributions. 
        The most stringent limits on CNC have been obtained as secondary results in experiments characterized by large masses and very low backgrounds~\cite{DAMA2000117301,BACK200229}.
        Therefore, the large detection efficiency, low threshold, and excellent energy resolution expected for CUPID are crucial to detect the low energy de-ectitation X-rays or Auger electrons. This, in combination with the ton-size scale of the experiment, makes us believe we can anticipate competitive results for these processes.



\subsection{Solar and Stellar Studies} \label{sec:SolarStellarStudies}
        \subsubsection{Sensitivity to Solar Neutrinos}
\label{sssec:SolarNeutrinos}
    Solar neutrino flux measurements have been an important probe for studying the inner mechanism of the Sun. In addition, it can provide valuable information to understand the properties of the neutrino itself, including neutrino oscillations. 
    \Mo isotopes are sensitive to electron neutrinos through CC interaction\cite{Ejiri:2000}; The electron neutrino interacts with a \Mo nucleus, excites it to the excited state of \Tc while emitting an electron and gamma rays. \Tc nucleus then decays to \Ru, emitting a beta particle and gamma rays. The combined energy of electrons and beta created in the process can be detected.  Calculations of the expected sensitivity of \cupidt{} and \cupid{} to the solar neutrino flux are underway.

    \subsubsection{Supernova neutrinos} \label{sssec:Supernovae} 
\begin{figure}
        \centering
        \includegraphics[height=2.3in]{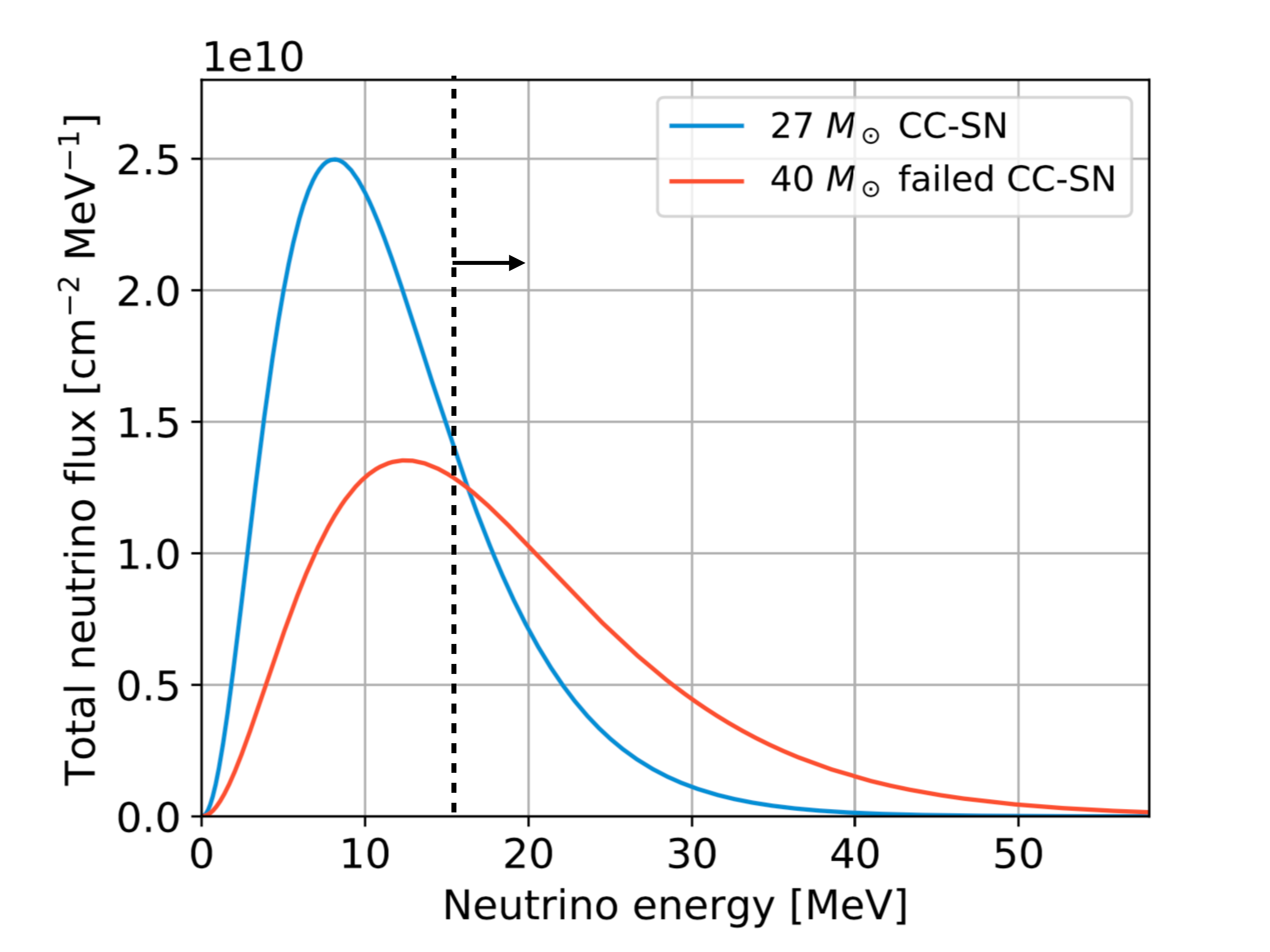}
        \includegraphics[height=2.3in]{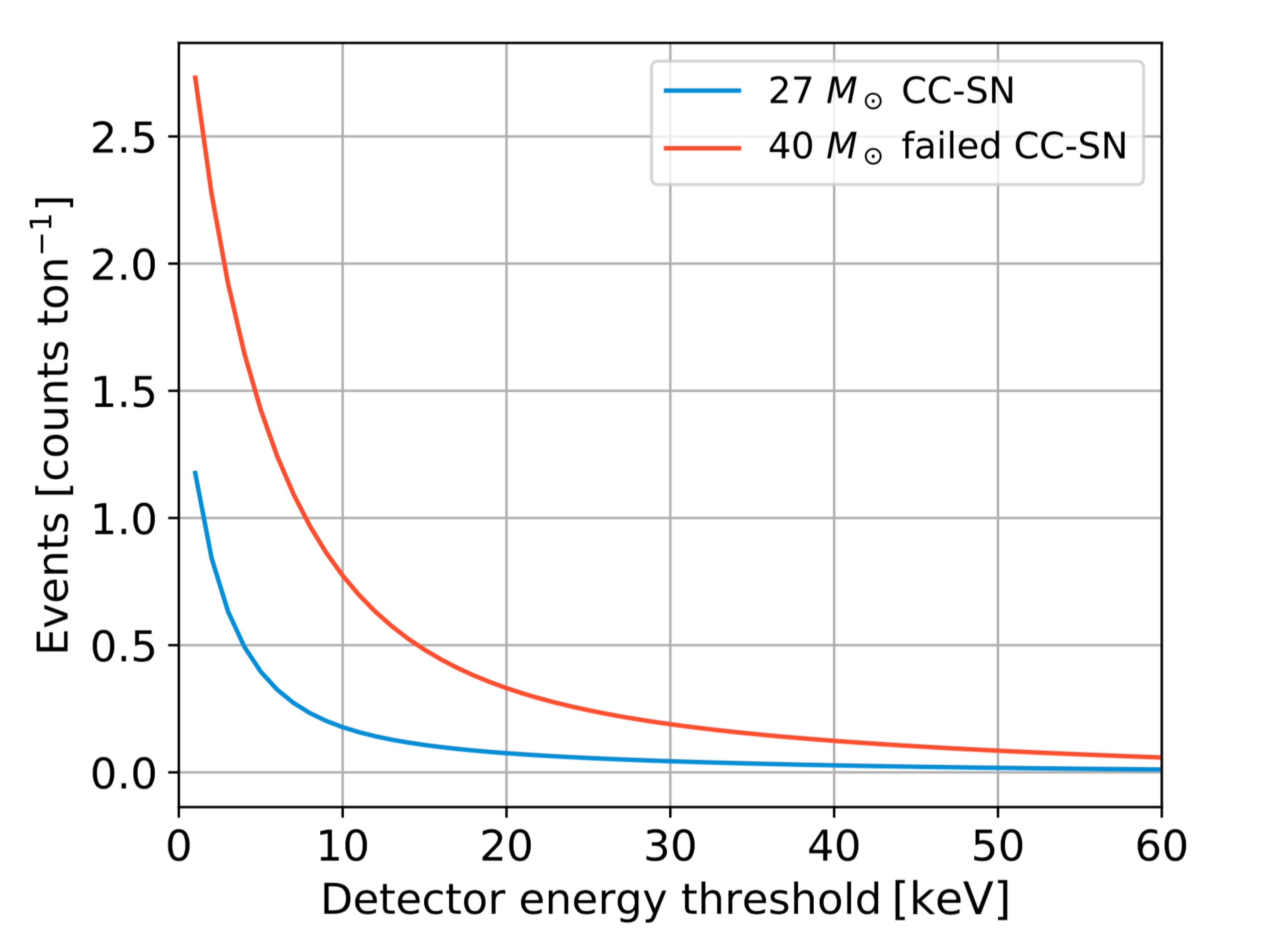}
        \caption{(\textit{Left}) Total neutrino energy spectra from two reference core-collapse SN events occurring at 10~kpc. The are on the right side of the vertical dashed line represents the portion of the neutrino energy spectrum detectable by CUPID. (\textit{Right}) Number of detected neutrino interaction in CUPID as a function of the detector energy threshold on the heat channel.} 
        \label{fig:SN_nu_EmissionAndThreshold}
    \end{figure}

    Supernovae (SNe) are among the most energetic events in the Universe, marking the end of a star's life. During a SN almost the entire binding energy of a star is released in the form of neutrinos/anti-neutrinos ($\nu_e/\overline\nu_e$ and $\nu_x/\overline\nu_x$ $x=\mu, \tau$), about 10$^{58}$ neutrinos of all flavors are shot out by the stellar explosion. Currently, little is known about the physics of core-collapse and the associated neutrino emission. Thus, detecting these elusive particles grants access to the processes and dynamics conspiring in this high energy event.
    Among the variety of detection channels available for the detection of astrophysical neutrinos, the one with the highest potential is coherent elastic neutrino-nucleus scattering (CE$\nu$NS)~\cite{Drukier:1983gj, Freedman:1973yd}. This neutral current process, being equally sensitive to all neutrino flavors, can enable the first full comprehensive detection of neutrino emission from SN. In addition, CE$\nu$NS has high cross-sections, about 10$^4$ times higher than other conventional neutral current processes (e.g. scattering on electrons) and no kinematic energy threshold, thus potentially sensitive to the full SN neutrino emission. The main challenge connected to the detection of SN neutrinos via CE$\nu$NS is the required nuclear recoil energy threshold for the detection of the neutrino scattering. However, as already demonstrated by various experiments exploiting the cryogenic calorimetric technique, ultra-low energy threshold can be achieved~\cite{Abdelhameed:2019hmk, Armengaud:2019kfj, Casali:2015gya, Fink:2020jts, Berge:2017nys}, and the calorimetric nature of the measurement enables a precise energy reconstruction without uncertainties caused by the energy quenching.
    In Fig.~\ref{fig:SN_nu_EmissionAndThreshold}, the time-integrated neutrino energy spectrum produced by a failed core-collapse SN with progenitor mass of $40\ M_\odot$ with fast accretion phase~\cite{Mirizzi:2015eza}, and a core-collapse SN with $27\ M_\odot$~\cite{Mirizzi:2015eza}, both occurring at a distance of 10~kpc. These neutrino energy distributions are the results of one-dimensional hydro-dynamical simulations carried out by the Garching group~\cite{garch}.

    The CUPID detector is expected to achieve an energy threshold on the heat channel of about 5~keV. Such low energy threshold enable the detection of SN neutrinos, however with limited sensitivity. In fact, as shown in Fig.~\ref{fig:SN_nu_EmissionAndThreshold} (left), CUPID will be able to detect only the higher energy tails of the neutrino energy distributions. CUPID's sensitivity to SN neutrinos can strongly benefit from a lower energy threshold.
    In order to compute how the CUPID detector response (i.e. number of detectable events) changes as a function of the detector energy threshold, we followed the prescription described in~\cite{Pattavina:2020cqc}. The results are shown in Fig.~\ref{fig:SN_nu_EmissionAndThreshold} (right).
    
\subsubsection{Solar axions} \label{sssec:SolarAxions} 

    Axions were first theorized as a solution to the Strong CP in Quantum Chromodynamics (QCD) by introducing an additional U(1) Peccei-Quinn (PQ) symmetry into QCD. The axion, a light, neutral, scalar or pseudoscalar boson, arises through the spontaneous symmetry breaking of the PQ symmetry \cite{PQ1977,Wilczek1978}. Axions' properties independently make them candidates for dark matter. Axion-Like Particles (ALPs) are a wider range of dark matter particle candidates that emerge from considering other broken symmetries. Axions and ALPs are well-motivated dark matter candidates, and it has been demonstrated that bolometer arrays have the potential to perform various solar axion searches through multiple production channels and detection methods \cite{DaweiLi2016, DaweiLi2015, CCVRAxions2013, Creswick1998, PRLSolarAxionsGe1998}. CUPID-1T would provide the highest exposure data set of low threshold, low backgrounds, high energy resolution, cryostat calorimeter data that could be utilized for solar axion searches to explore potential beyond the Standard Model physics. 
    
    The Sun could be a rich source of axion production, creating an axion flux to Earth through various mechanisms. Based on the standard solar model, a solar axion flux could be produced by the Primakoff conversion of blackbody photons to axions in the Sun. This would create a solar axion spectrum that peaks at 3 keV \cite{VanBibber1989}. This channel is sensitive to the coupling constant $g_{a\gamma}$. The Sun could also produce a flux of mono-energetic axions through Atomic-Bremsstrahlung-Compton (ABC) production channels such as the 6.4 keV X-ray de-excitation peak from Fe \cite{Redondo2013}. Thermal de-excitations from the M1 nuclear ground-state transition, such as the 14.4 keV and 477.6 keV excited states of ${}^{57}$Fe and ${}^7$Li, respectively could additionally produce a mono-energetic axion signals sensitive to the coupling constant $g_{an}^{eff}$ \cite{DaweiLi2016,Krcmar2001}.
    
\begin{figure}[h]
    \centering
        \includegraphics[width=0.7\textwidth]{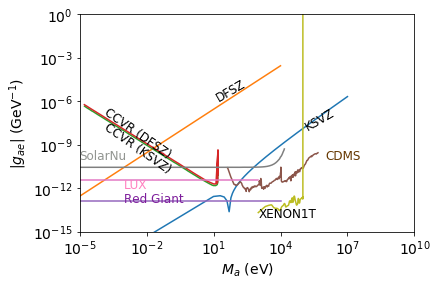}
    \caption{$g_{ae}$ and $m_a$ limits for CCVR solar axion search assuming a DFSZ and KSVZ axion are shown \cite{CCVRAxions2013} along with limits for Red Giants \cite{RedGiantsPhysRevD.102.083007}, CDMS \cite{CDMSPhysRevD.101.052008}, LUX \cite{LUXPhysRevLett.118.261301}
    Solar Neutrinos \cite{SolarNuPhysRevD.79.107301}, and XENON 1Ton \cite{XENON1T1PhysRevLett.123.251801}
        \cite{XENON1T2PhysRevD.102.072004}. 
        Data and plot was adapted from Reference \cite{AxionWebsiteciaran_o_hare_2020_3932430}
        }
     \label{fig:Gae_CUORE}
\end{figure}

    CUPID's predecessor, CUORE has focused on two potential detection mechanisms for solar axions in TeO$_2$ crystals, which would also be applicable to CUPID and CUPID 1T. The first mechanism is the axio-electric effect, where axions could be absorbed in a crystals. This mechanism relates to $g_{ae}$, and it relies on axions having the ability to couple to electrons. A search for solar axions using this technique to detect 14.4 keV solar axions was completed with data from a CUORE Crystal Validation Run (CCVR) of 4 crystals, with a total exposure of 43.65 kg·d of data \cite{CCVRAxions2013}. The results of this search compared to other axion limits are shown in Figure \ref{fig:Gae_CUORE} \cite{CCVRAxions2013}. 
    
\begin{figure}[h]
\centering
   \includegraphics[width=0.7\textwidth]{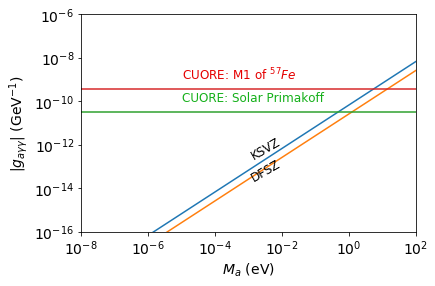}
   \caption{Sensitivity to $|g_{a\gamma\gamma}|$ and $m_A$ for the QCD Axion (KSVZ and DFSZ) \cite{PDGPhysRevD.98.030001} in the M1 transition of $^{57}Fe$ \cite{DaweiLi2016} and the Solar Primakoff spectrum detected through inverse coherent Bragg-Primakoff conversion technique in CUORE \cite{DaweiLi2015}.}
   \label{fig:Gayy_CUORE}
\end{figure}

    Secondly, sensitive estimates have been performed for the Inverse Coherent Bragg-Primakoff Conversion technique \cite{Creswick1998, PRLSolarAxionsGe1998,DaweiLi2016,DaweiLi2015}. In this mechanism, axions would couple to charges in a crystal by a virtual photon of the Coulomb field. Axions convert to photons when the axion incident angle satisfies a Bragg condition determined by the crystalline structure of the bolometer. This produces a time-dependent signal based on the relative position of the sun and the crystal axes. The Coherent Bragg-Primakoff mechanism is sensitive to $g_{a\gamma}$, and the sensitivity of the CUORE experiment using this mechanism to search for 14.4 keV solar axions and the spectrum of solar Primakoff axions is shown in Figure \ref{fig:Gayy_CUORE}. Barriers to sensitivity and ongoing searches in CUORE include reducing analysis thresholds below 5 keV as well as understanding low energy backgrounds.
    
    With the appropriate measure of crystal structure and installation as well as lowering analysis threshold and backgrounds, CUPID 1-T will be sensitive to these same channels.  Sensitivity studies of CUPID and CUPID-1T  are ongoing, but with a high exposure CUPID-1T is expected to reach new regions of parameter space for a lab-based crystal calorimeter array.

    
    Resonant absorption of axions in ${}^7$Li is an additional detection mechanism available to CUPID-1T. In this mechanism, an axion would be produced by the 477.6 keV M1 nuclear transition of excited ${}^7$Li states in the Sun \cite{Moriyama1995}. ${}^7$Li atoms in Li$_2$MoO$_4$ crystals in CUPID-1T could absorb these axions to form excited states, and produce a 477.6 keV signal from de-excitation \cite{SNOSolarAxions2021,Belli2012}. This method is sensitive to $g_{aN}$. This mechanism produces a higher energy signal than the other discussed solar axion channels, which may enable a search with lower backgrounds and unrestrained low thresholds requirements in bolometers. With 1871 kg of Li$_2$MoO$_4$, crystals, CUPID-1T will contain 147.6 kg of Li. 
    


In addition to searches for symmetry violations and new decay modes, \cupidt is expected to have sensitivity to products of a number of stellar processes.  Sensitivity to solar and supernova neutrino fluxes may reveal new information about neutrino properties and stellar astrophysics,  while searches for solar axions may result in new bounds or new avenues of discovery for beyond standard model physics.

\subsection{Dark matter \& Additional Topics} \label{sec:DMMiscBSM}

The search for beyond-\sm physics with \cupidt also holds great potential in the arena of dark matter searches, where large exposures, low backgrounds, low energy thresholds, and sharp resolutions play a tremendous role in detector sensitivity.  In addition, the ability to study multiple isotopes simultaneously and a segmented detector structure will allow \cupidt to perform interesting searches for several different prospective signals, and which will contribute to a search strategy in many ways complementary to current and future direct searches for dark matter and other new particle interactions. 

 \subsubsection{ Low-Mass Dark Matter} \label{sec:LowMassDarkMatter} 
        
Although its existence is strongly suggested by several gravitational effects~\cite{DELPOPOLO20131548}, dark matter (DM) represents one of the deepest mysteries in modern physics and its nature still remains unknown. 
While many candidates for dark matter exist, a few classes of candidates, such as Weakly Interacting Massive Particles (WIMPs), axions, and Majorons (See section \ref{sec:Majorons}) have received particular notice, being both theoretically motivated and well suited to current search strategies \cite{pdgdm:2021}.  The WIMP hypothesis is one of the longest trending searches and remains one of the most attractive and simplest scenarios to explain DM, but the absence of an observed signal has led to very stringent constraints on WIMP properties, encouraging physicists to look for alternatives and stronger signatures~\cite{ROSZKOWSKI2018066201,ARCADI2018203}. 
In this framework, the possibility of observing a time-varying signal related to the relative motion of the earth with respect to the DM halo of the galaxy has attracted considerable interest~\cite{MAYET20161}.
A low energy threshold and a large mass are the most appealing features for experiments hoping to observe these effects.  With the possibility of separating nuclear recoils (characterized by a lower light yield) from the dominant electron-based background, CUPID's technology is particularly well-suited for searches of this kind.  We are exploring the sensitivity for observing the WIMP seasonal modulation generated by the motion of the Earth. 

\subsubsection{Magnetic monopoles (MM) searches}
\label{sec:MM}

Magnetic monopoles (MM) were introduced by P. Dirac in 1931~\cite{Dirac:1931kp} to explain the discrete nature of the electric charge. These particles, yet to be detected experimentally, have magnetic charge and are allowed by Maxwell's equations while keeping their consistency and without being in contradiction with any experimental results. Theories beyond the Standard Model, for example Grand Unified Theories (GUTs), predict monopoles\cite{tHooft:1974kcl}.

\begin{figure}[htp!]
        \centering
        \includegraphics[width=0.5\textwidth]{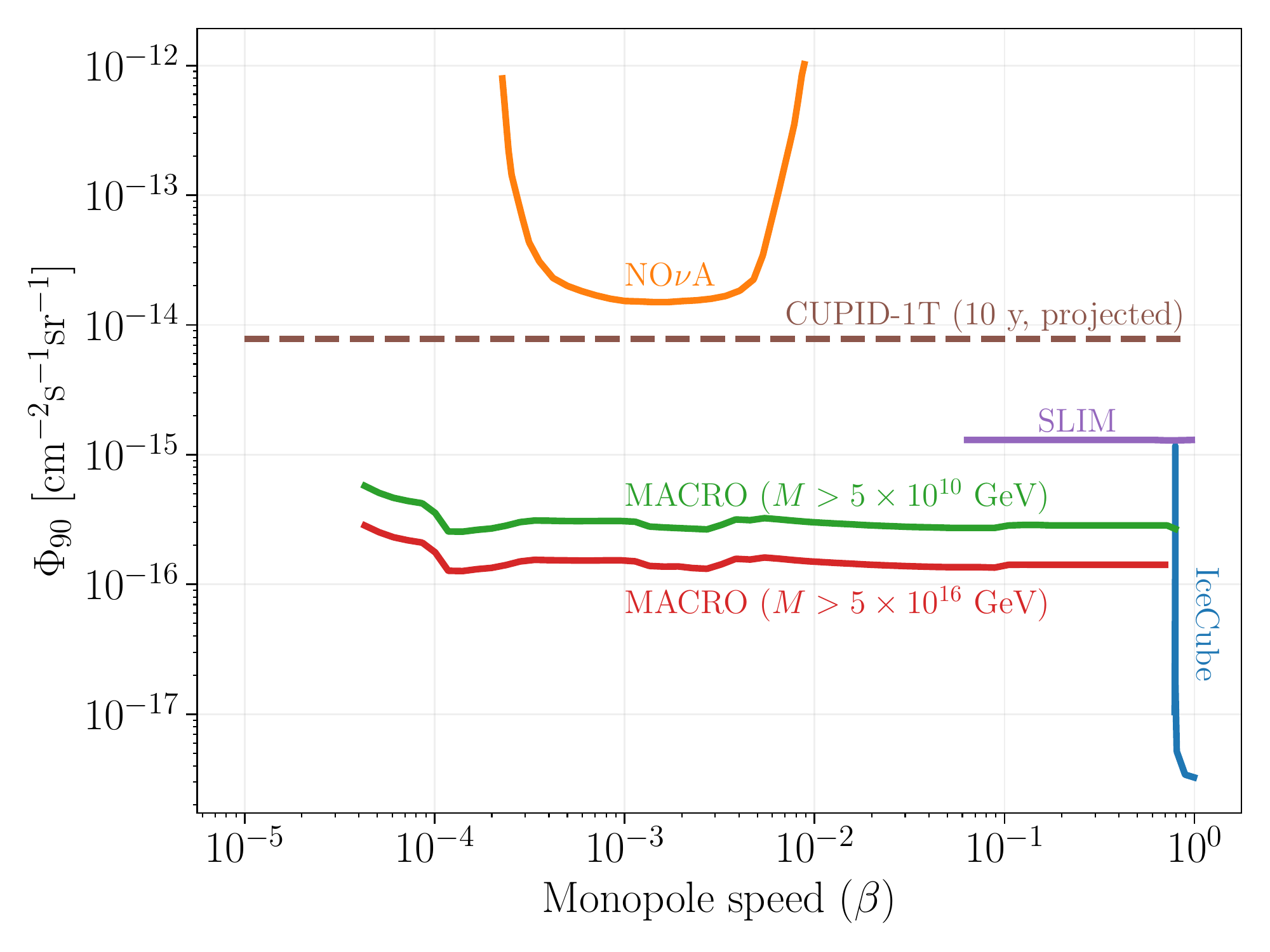}
        \caption{Upper limits on the isotropic flux of magnetic monopoles as a function of $\beta$ from the NO$\nu$A~\cite{NOvA:2020qpg}, MACRO~\cite{MACRO:2002jdv}, SLIM~\cite{SLIM:2008ps}, and IceCube~\cite{IceCube:2014xnp} experiments. Also shown the projected single-event sensitivity that CUPID-1T would have with a livetime of 10 years.}
        \label{fig:monopoles_sensitivity}
\end{figure}

When passing through matter, MM lose energy at a rate that depends strongly on their velocity ($\beta$)~\cite{Cecchini:2016vrw}. The signature property of fast monopoles ($\beta>10^{-2}$) is their large energy loss rate compared slower monopoles or to minimum ionizing particles such as muons~\cite{Cecchini:2016vrw}. Slower monopoles, on the other hand, could loss energy at rates comparable to those of muons, making the latter the principal background for these searches. CUPID will be equipped with a muon-tagger, with a proposed time-of-flight resolution of the order of nanoseconds, that can reject muon events with an efficiency $>99\%$, making CUPID sensitive to fast and slow monopoles ($\beta \in [10^{-4},1]$). Additionally, CUPID's thermal readout will be sensitive to all energy depositions allowing the exploration of the parameter space at very-low $\beta$. Fig.~\ref{fig:monopoles_sensitivity} shows existing limits from different experiments to the magnetic monopole flux, as well as the projected single-event sensitivity that CUPID-1T could achieve with a livetime of 10 years.

\subsubsection{Track-like events} \label{sec:TracklikeEventReconstruction}

Benefiting from its significant exposure and high granularity, CUPID will be able to serve as a tracking calorimeter for searches for exotic cosmic particles crossing the entire detector volume.  Dedicated tracking algorithms can leverage the segmented geometry of the detector to reconstruct detector-wide events\cite{yocum2022muon}, while the addition of an external muon veto detector system can provide additional time-of-flight information to separate relativistic and non-relativistic particles. 

In particular, we consider below two classes of Beyond-the-Standard-Model particles that could be searched for with CUPID implemented as either a single or multi-location experiment, namely magnetic monopoles and lightly ionizing particles (LIPs). Other BSM phenomenology such as Multiply Interacting Massive Particles (MIMPs)\cite{PhysRevD.98.083516} are also expected to leave track-like signatures, and could be searched for by CUPID.  Additionally, the bolometric technique employed within CUPID is sensitive to all manner of energy depositions, so would still be sensitive to novel energy-loss interactions, or ultra-heavy particles for which individual energy depositions are below ionization threshold, but still contribute thermal energy into the detector.

\subsubsection{ Lightly-ionizing (LIPs), and fractionally charged particles}
\label{sec:lips}
Free and stable particles with fractional electric charge arise in different BSM scenarios as cosmic relics including millicharged dark matter \cite{PhysRevLett.110.241304}, and have gained attention at collider-based searches for new physics as well \cite{Pinfold:2019zwp}.  Such lightly ionizing particles will have reduced stopping power within matter, leading to track-like energy depositions within detectors below that of minimally ionizing particles with unit charge.

\begin{figure}
\centering
\includegraphics[width=4.0in]{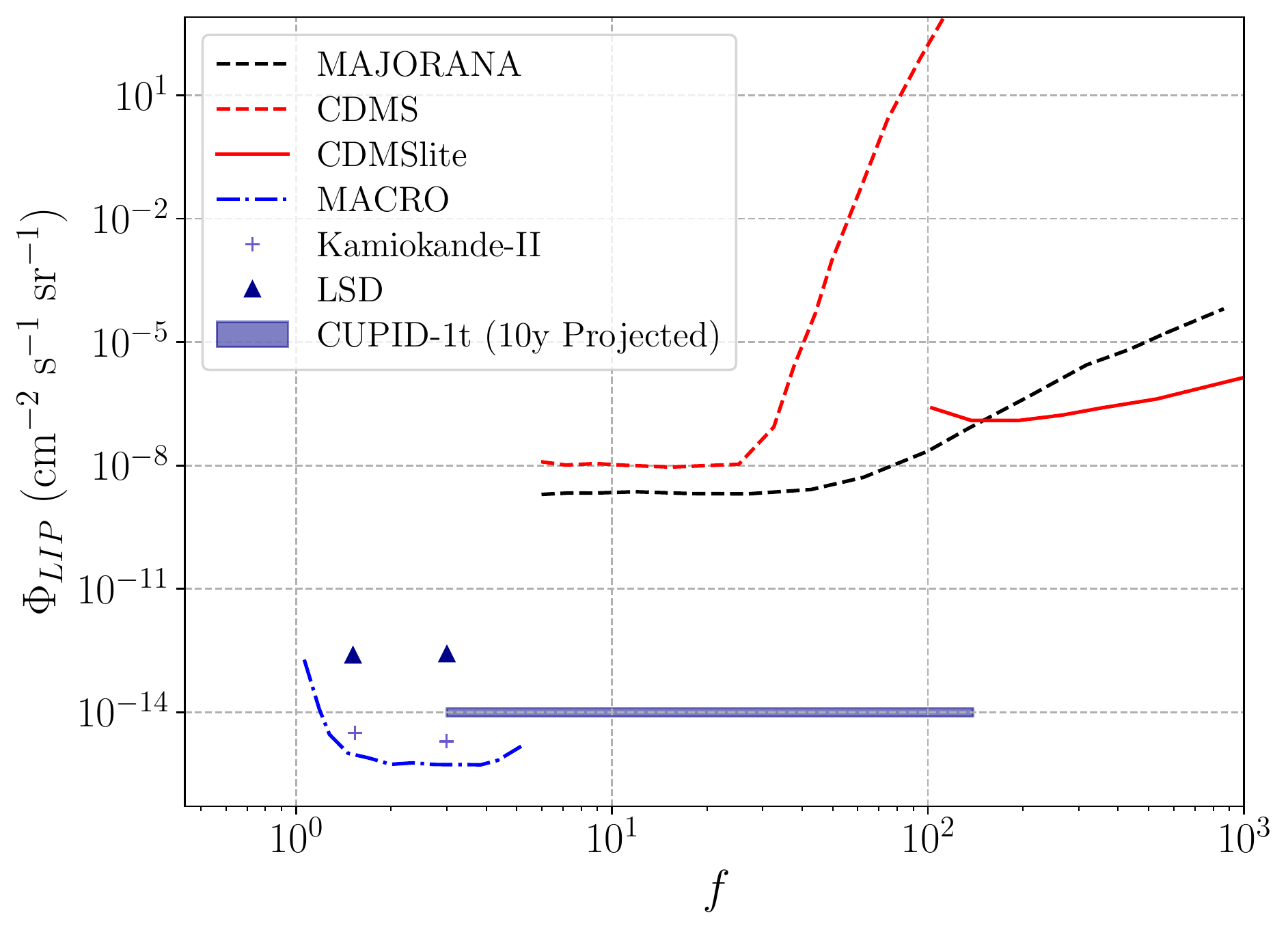}
\caption{Projected sensitivity of CUPID-1t to Lightly Ionizing Particles for 10 years of livetime, compared with past search limits\cite{PhysRevLett.120.211804,AGLIETTA199429,PhysRevD.43.2843,PhysRevLett.114.111302,PhysRevD.62.052003}.}
\label{fig:cupid-lip-sensitivity}
\end{figure}

CUPID's high energy resolution, low trigger thresholds and high granularity make it excellent for searching for LIPs. Parameterizing the LIP charge as $q=e/f$, Figure~\ref{fig:cupid-lip-sensitivity} shows the expected LIP sensitivity of CUPID-1T with 10 years of exposure, covering much of the parameter space between millicharged searches at high-$f$ and large underground tracking detectors. 
        
\section{Longer-term R\&D on Advanced Detector Technologies}
\label{sec:randd}
The physics goals of the \cupidt experiment must be supported by a robust program of research and development of the required technology.
        CUPID groups are actively involved in several key efforts to support the dual goals of multiplexed readout and background reduction. Areas of particular interest for \cupidt{} include the use of high-speed superconducting sensors 
        for thermal~\cite{Wang:2015taa} or athermal~\cite{Alkhatib:2020slm} phonon detection; the adaptation of multiplexed readout technologies (synergy with CMB) to macrobolometers
        ; single-crystal topological reconstruction through multi-sensor phonon and photon imaging; the development of an active $\gamma$ veto (in synergy with low-mass dark matter experiments)
        ; and the incorporation of CMOS and ASIC developments for quantum sensors (in synergy with CMB, DM, and \qis{}) \cite{Huang:2020hrv}.  All of these efforts, as well as international R\&D on the use of superconducting crystal coatings to enhance PSD capabilities (including work by \cross{} at the Canfranc underground laboratory \cite{Zolotarova:2020suz}) have the potential to profoundly impact the design and fabrication of bolometric detectors for fundamental science.
        

\subsection{Superconducting Sensors}\label{TES_MKID}
        
        Neutron-transmutation doped thermistors (NTDs) are expected as part of the baseline design for CUPID\cite{cupid2019}, but requirements for fast rise time and limits on the \dbd pile-up background motivate the development of other types of sensors. Multiple modes of superconducting sensors are under development as we look toward CUPID-1T: Microwave Kinetic Inductance Detectors (MKIDs), Metallic Magnetic Calorimeters (MMCs), and high- and low-impedance Transition Edge Sensors (TESes). A selection is highlighted here. 
        
        \subsubsection{MKIDs}
        \label{subsec:mkids}
                Microwave Kinetic Inductance Detectors (MKIDs\cite{Day:2013,Zmuidzinas_2012}) are one of the most promising technologies in the field of astrophysics\cite{mazin:2010,Adam:2017gba}. 
An MKID consists of a superconducting $LC$ circuit operated with an alternating current, and acting as a resonator. The resonant frequency, $f_{\mathrm{res}}$, can be tuned by adjusting the values of $L$ and $C$ ($f_{\mathrm{res}}=\frac{1}{2\pi\sqrt{LC}}$) and is usually in the range of GHz. With small variations of $L$ or $C$, it is possible to design hundreds of MKIDs resonating at slightly different frequencies, and to read them out using a single cable (frequency multiplexing).
The absorption of energy in the superconductor breaks Cooper pairs into quasiparticles, changing the value of the inductance and, as a consequence, of $f_{\mathrm{res}}$. However, the quasiparticles are dissipative, which results in a broadening of the resonance peak.
By monitoring variations in the signal transmitted past the MKID, it is possible to reconstruct the initial energy deposit. 

Experiments like CUPID and \cupidt (or like the AMoRE experiment \qtd{\cite{Lee_2020}}) 
demand thousands of sensitive and reproducible light detectors to be operated at 10\,mK. The ease in fabrication, high sensitivity, and natural frequency multiplexing offered by MKIDs are particularly appealing to satisfy such constraints. 
For this reason, the CALDER project \cite{cardani_calder2021} was designed to port the MKIDs technology to the field of particle physics. The project designed novel light detectors by following a ``phonon-mediated" approach\cite{Moore:2012au, Battistelli:2015vha}: photons emitted by the cryogenic calorimeter are absorbed into an insulating substrate and converted into phonons. Phonons travel in the substrate itself until they are absorbed by the MKID, resulting in a variation of the transmission past the device. While this approach comes at the cost of a low efficiency (photons have to be converted into phonons rather then being directly absorbed in the MKID), it provides coverage of wide detector areas (several cm$^2$) using a limited number of devices, thus simplifying the readout.
 
The CALDER project began with a 2$\times$2\,cm$^2$ substrate equipped with four aluminum MKIDs \cite{cardani:APL2015}.  Extensive R$\&$D on the MKID geometry\cite{cardani:APL2017} and material\cite{Cardani_SUST_2018} resulted in significant improvements in detector performance, achieving a noise RMS of 26\,eV using a single MKID made of a multi-layer aluminum-titanium-aluminum deposited on a 2$\times$2\,cm$^2$ substrate.
The high sensitivity of the prototypes, as well as their fast time response, enabled the first measurement of the scintillation time-constant of a Li$_2$MoO$_4$ crystal at 10\,mK\cite{Casali_2019}. This was an important milestone to fully explore the potential of the idea of rejecting pile-up events in Li$_2$MoO$_4$ using the (faster) light signals instead of the calorimetric ones.
 
In the second phase of the project, the area of the light detector was increased to 5$\times$5\,cm$^2$,  the original requirement for the CUPID experiment. The CALDER collaboration demonstrated a noise RMS of 34\,eV and a rise-time of 120\,$\mu$s\cite{cardani_calder2021}, successfully concluding the R$\&$D on the detector. 
The exploitation of these devices in a next-generation project would require efforts in (i) integrating such devices in a CUPID-like detector array and (ii) interfacing the MKID multiplexed readout with the NTD readout. Commercial readout boards such as the model N321 from Ettus Research \cite{ettus} complemented by open source software for the data acquisition\cite{minutolo} already meet these requirements.
        \subsubsection{Low impedance TES}
                Low impedance transition-edge sensors (TESs) have been widely used and play a pivotal role in astronomy and astrophysics instrumentation\cite{ullom2015review,gottardi2021review}. The sensors are biased in their superconducting-to-normal transition, and a calorimetric measurement is done by measuring the changes in resistance due to energy-induced temperature excursions. The sensitivity of a sensor is typically defined by the quantity $\alpha \coloneqq (T/R)(dR/dT)$ which signifies the logarithmic dependence of the resistance on temperature excursion. One can achieve a high value of $\alpha$ --- $\mathcal{O}(\sim$ 100--1000) --- for a TES using a suitable superconducting film. In contrast, semiconductor thermistors typically have $\alpha$'s of $\mathcal{O}({\sim}1{-}10)$\cite{stahl2005cryogenic}. Therefore, the TESs offer superior speed and potentially better resolution than a semiconductor thermistor. Moreover, there is a lack of suitable multiplexing schemes for semiconductor thermistors. On the other hand, multiplexing schemes for TESs for $\mathcal{O}(> 1000)$ of TESs have been demonstrated in the field of astronomy. Owing to the factors listed above, the TESs remain a compelling technology for light-detector sensors for CUPID-baseline, and more so for the CUPID-1T experiment where multiplexing will be necessary.

US-led groups are actively involved in developing low $\mathrm{T_c}$-TES-based light-detectors. US-led groups have developed novel proximity-coupled Iridium-Platinum bi-layer (and Iridium-Gold tri-layer) superconducting films, and demonstrated tuning of the critical temperature ($\mathrm{T_c}$) down to 20~mK by varying the thickness of the normal metal (Pt or Au)\cite{hennings2020controlling}. The first generation of TES-based light detectors uses a sputter-deposited  ($\mathrm{300~\mu m \times 300~\mu m }$) Ir-Pt ($\mathrm{t= 100~nm/60~nm} $) film on top a intrinsically pure Silicon substrate ($\mathrm{\oslash = 50.8~mm; t = 285~\mu m}$). Si was chosen because of its radiopurity, low cost, and higher Debye temperature ($\mathrm{\theta_D\sim636~K}$)\cite{LandoltBornstein2002:sm_lbs_978-3-540-31356-4_478} compared to Germanium~($\mathrm{\theta_D\sim374~K}$)\cite{LandoltBornstein2002:sm_lbs_978-3-540-31356-4_532}. The sensor's small size and high Debye temperature of the photon absorbing substrate are ideal for getting a low heat capacity for the detector and increasing its sensitivity. The superconducting film is contacted with Nb leads, --- which have much higher $\mathrm{T_c}$ and critical current than the bi-layer-TES --- and placed in parallel with a small shunt resistance, which allows for voltage-biasing of TES. A voltage-biased TES utilizes electro-thermal feedback to prevent thermal runaway\cite{irwin1995application}. Subsequently, we use a superconducting quantum interference device (SQUID) ammeter to read out the TES due to its low noise, low impedance, and high bandwidth. 

\begin{figure}
    \centering
    \includegraphics[width=0.40\textwidth]{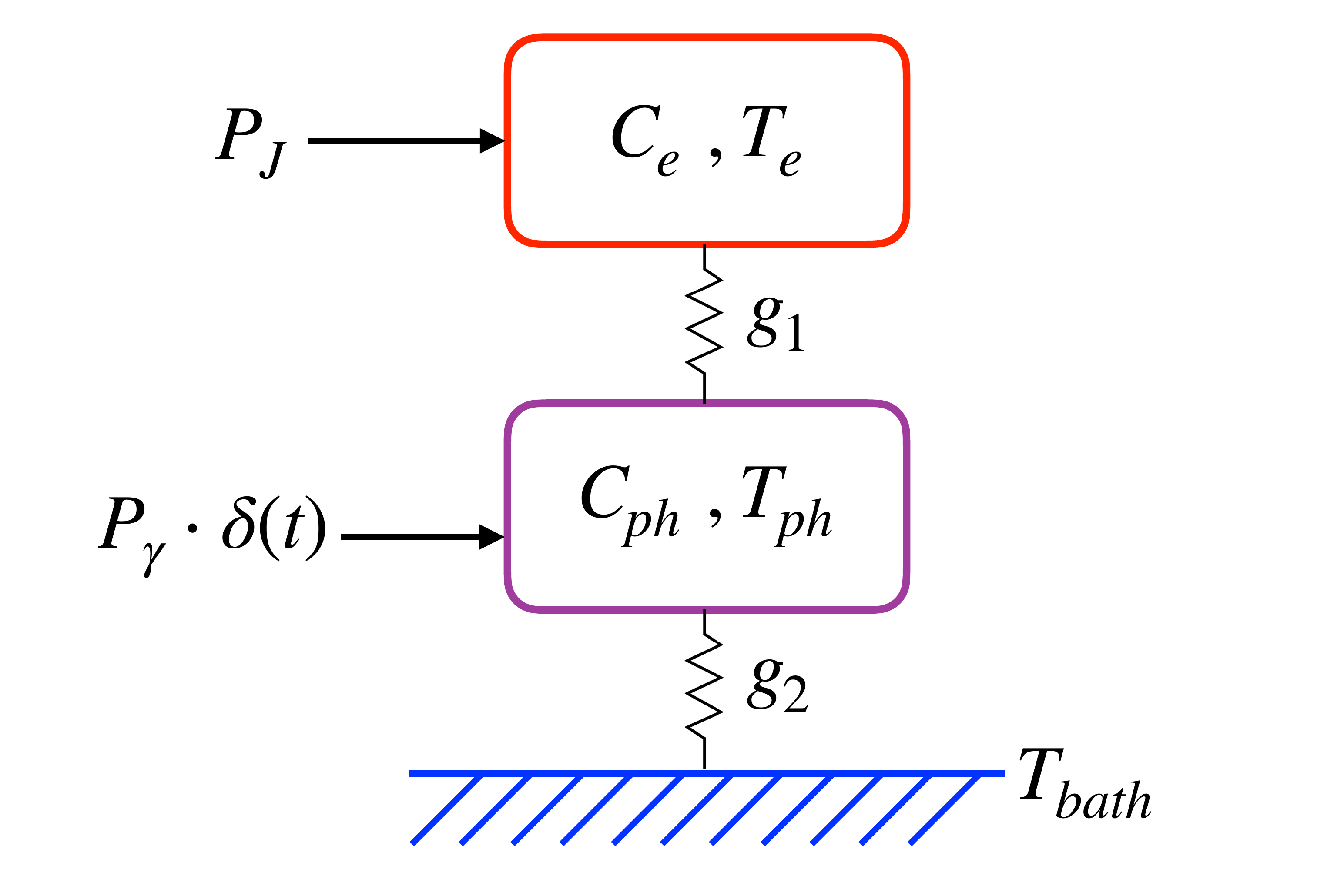}
    \caption{A simplified schematic of two-block model for thermal flow in TES-based light-detectors. See text for details.}
    \label{fig:two_block_model}
\end{figure}

\begin{figure}
     \centering
     \begin{subfigure}[b]{0.48\linewidth}
         \centering
         \includegraphics[width=1.19\textwidth]{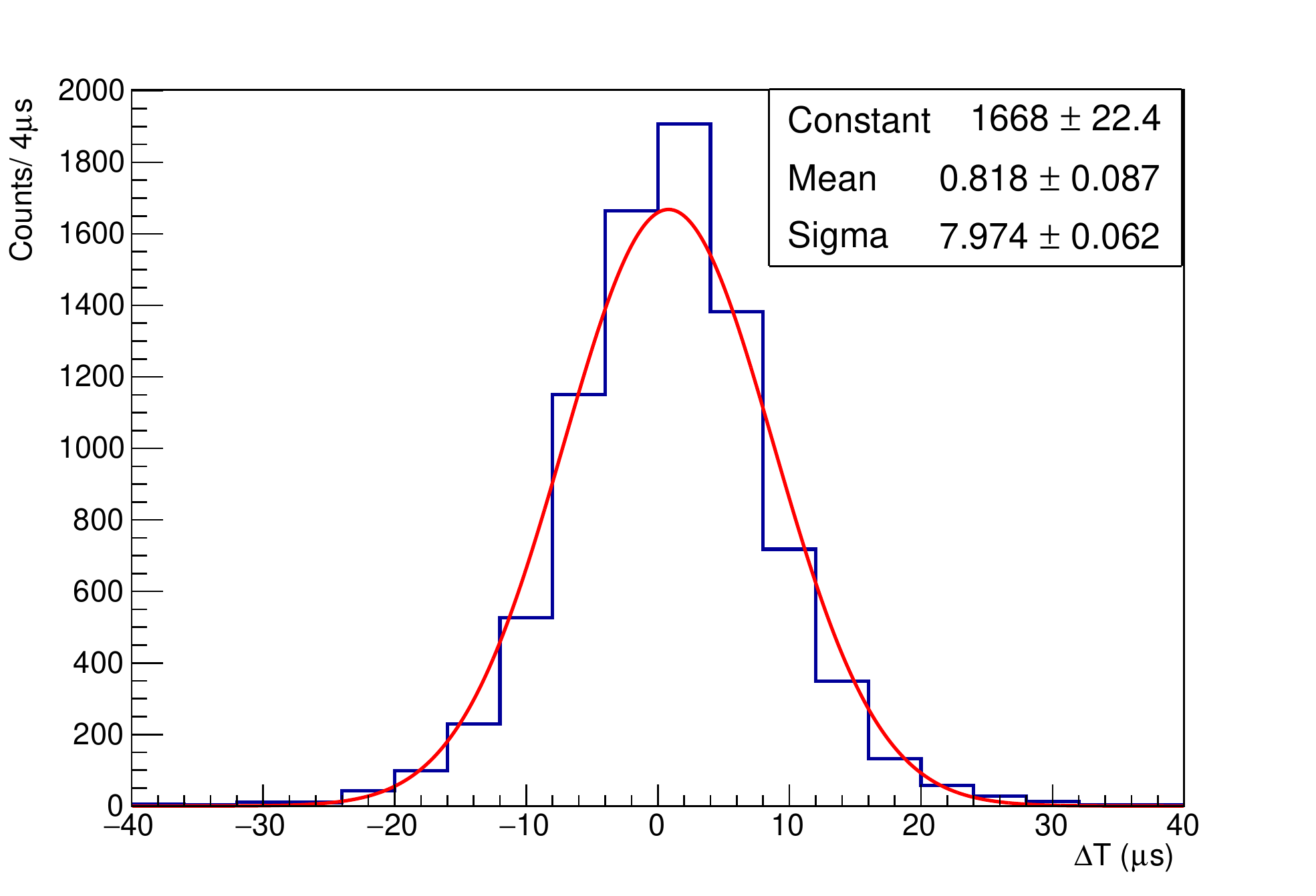}
         \caption{Distribution of difference in trigger times of two consecutive LED pulses. We demonstrate the timing resolution of $\sim$~8 $\mu$s.}
         \label{fig:TR_LD}
     \end{subfigure}
     \hfill
     \begin{subfigure}[b]{0.48\linewidth}
         \centering
         \includegraphics[width=\textwidth]{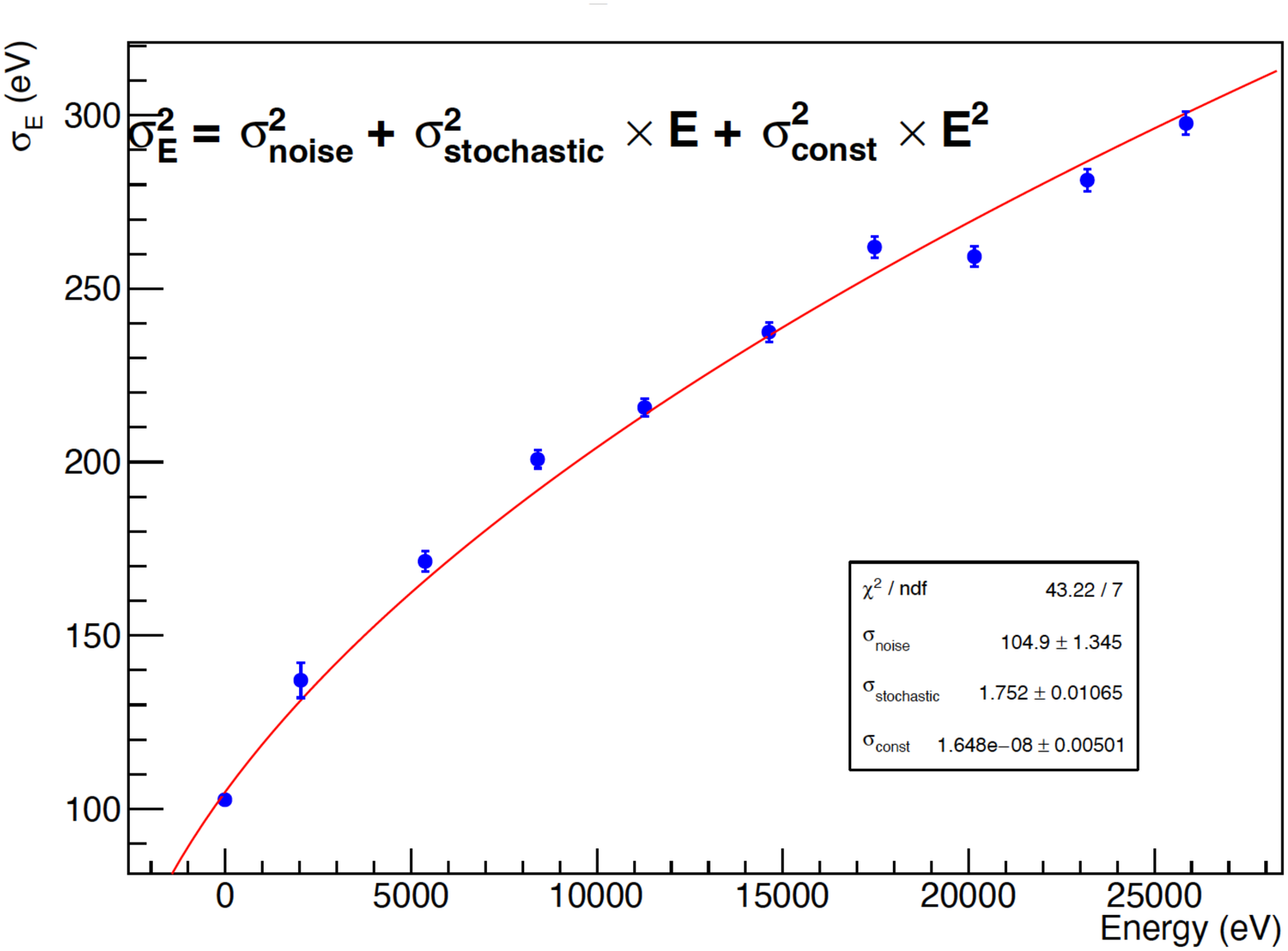}
         \caption{Measured $\sigma_{baseline}$ $\sim$ 100~eV. The detector is calibrated using Poisson statistics of photon pulses from light-emitting diodes outside the cryostat.}
         \label{fig:ER_LD}
     \end{subfigure}
        \caption{Timing and energy resolution of TES-based prototype light detector. We demonstrate that this detector meets the CUPID requirements. }
        \label{fig:TES plots}
\end{figure}

We have demonstrated excellent timing resolution of $<$10~$\mu$s and a baseline energy resolution of $\sim$100~eV on our prototype TES-based light detectors (Fig.~\ref{fig:TES plots}) in a pulsed-tube based cryostat. The timing resolution is an order of magnitude better than that of the NTD-Ge based light detectors. Furthermore, the achieved baseline resolution is on par with the performance of NTD-Ge based light detectors. This is a significant accomplishment for detectors of this size that use fabrication technology that easily scales to 1000s of devices. However, there are several technical challenges that still need to be overcome. The device physics at the microscopic level is still relatively unknown. Phonon loses due to escape of ballistic phonons, creation of long lived excitations in insulators, and fluctuation in created electron-hole pairs are some of the reasons due to which we can observe degradation in energy resolution\cite{kozorezov2013athermal}. 
Moreover, there is scope for optimizing the phonon transport in the detector by engineering the acoustic mismatches at different interfaces.  

Despite these challenges, the TES-based technology with an appropriate multiplexing scheme will, perhaps, be indispensable for the CUPID-1T experiment where the required rejection of the $2\nu\beta\beta$ pile-up background is impossible with NTD-Ge sensors. 
If ready in time, the TES technology and its multiplexed readout could also be made available for the CUPID-baseline experiment, improving its sensitivity. 

\subsection{Developments in Cryogenic and Experimental Design Beneficial to Quantum Sensing Technology} \label{sec:CMOS_ASIC}
        
        All electronics used to bias CUORE \& CUPID's NTD sensors are operated at room temperature. This scheme offers operational simplicity and ease of maintenance, but comes with a number of drawbacks. In particular, this means that the NTD signals must be sent through meters of wire from the 10~mK stage to room temperature before any preamplification. This long cable length exposes the system to electronic and vibrational noise pickup, and the associated cable capacitance introduces latency that can spoil pulse shape fidelity that will be important to the more stringent pileup rejection demands of CUPID-1T (see Section \ref{sec:CUPID1T_Requirements}). 
Reducing the number of wires exiting the cryostat is a firm requirement for CUPID-1T, as discussed further in Section~\ref{sec:Multiplexing}.


Conventional CMOS-based ASIC devices operated at temperatures less than 4~K may present an alternative approach for signal amplification. By placing electronics at cryogenic stages in close physical proximity to our sensors, we can minimize pickup noise and multiplex the system to reduce the wire load exiting the cryostat. In addition, the manufacture of such devices is easily scalable to the numbers needed for thousands to tens of thousands of sensors in CUPID-1T, as we can take advantage of the maturity of industrial semiconductor processes. The idea of using conventional CMOS electronics at cryogenic temperatures has already attracted interest in the quantum computing community, which faces similar problems of needing high-fidelity control and readout of large numbers of qubits that are placed at the coldest stages of dilution refrigerators to avoid interference from thermal fluctuations \cite{CryoCMOSQC1, CryoCMOSQC2, QCVibNoise}. The development of cryogenic CMOS-based ASICs for TES and NTD readout in CUPID-1T thus offers additional synergy with the needs of quantum computing (see Section \ref{sec:BroaderImpacts_QuantumComputing}). 

\begin{figure}
    \centering
    \includegraphics[width=0.5\textwidth]{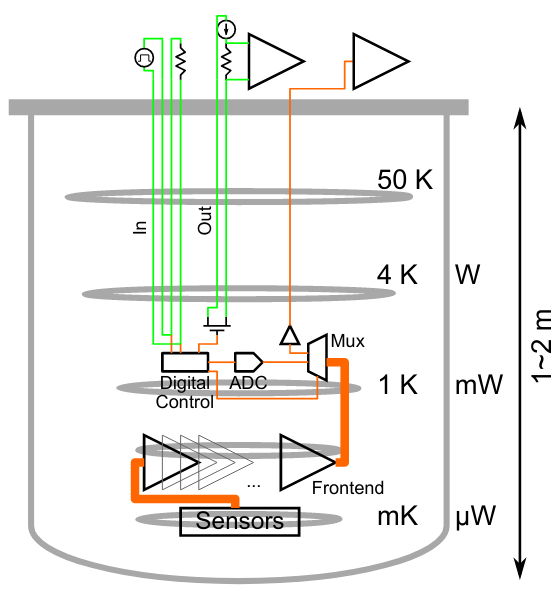}
    \caption{Schematic design of cryogenic electronics integrated into CUPID. To minimize their distance to sensors, front-end preamplifiers should be located below the still. Supporting electronics may be placed at slightly higher stages, where more cooling power is available, and from where they can drive the signal up the rest of the cable length to outside the cryostat. (Figure 10.1 in \cite{Huang:2021})}
    \label{fig:CMOSinCUPID}
\end{figure}

An example of the potential incorporation of cryogenic electronics into CUPID is shown in Fig. \ref{fig:CMOSinCUPID}. In this scenario, the front-end preamplifiers are located as close to the sensors as possible, below the still of the cryostat. Other supporting electronics may be placed slightly higher, at the 1~K or even 4~K stage where power constraints are much looser.  This allows more freedom in design while maintaining the benefits of placing the electronics at the cold stages of the cryostat.

NMOS and PMOS transistors manufactured using the TSMC 180-nm process have now been successfully operated and characterized at temperatures down to less than 100 mK, showing qualitatively similar behavior to that seen at 4 K \cite{Huang:2020hrv, Huang:2020}. Building upon this work, circuits using the CMOS 180-nm process have also been tested in this deep cryogenic regime \cite{Huang:2021,Huang:2021a}. These studies show that conventional CMOS electronics can indeed be pushed down to the mK stages of the CUPID-cryostat without special modifications, and they have set the groundwork for showing that cryogenic CMOS-based ASICs have the potential to satisfy both the signal requirements and power constraints of a CUPID-style experiment.



\subsection{Multiplexed Readout Technologies}\label{sec:Multiplexing}
         

The increase in mass for a next-to-next-generation crystalline detector already projects tens of thousands of channels, which requires advanced technical solutions to read out. The additional instrumentation for multi-mode signals requires advanced multiplexing solutions to readout channels (up to order twelve sensors for each crystal, for ongoing demonstrators, see Section \ref{sec:Synergies_DEMETER_MUX}); this easily compounds into tens of thousands more channels to be read out simultaneously. Multiplexed readout is a critical juncture between creating a background-free search for lepton number violation and advancing developments in the field of quantum information science.

While MKIDs (see Section \ref{subsec:mkids}) are one natural choice for a multiplexed array, the readout of transition edge sensor (TES) arrays at the level of ten thousand channels has been demonstrated using several multiplexing technologies. Time-division multiplexing (TDM) — a technique by which synchronized switches control rapid access to multiple channels in sequence — requires thousands of wires from the cryogenic stage out to room temperature electronics. Efforts to decrease the wire density in this technique lead to degradation of noise performance, driving the need for improved technology. Alternatively, frequency-domain multiplexing (FDM) uses superconducting resonators to address many TESs simultaneously. Because FDM does not use switching between channels, it requires far fewer cryogenic cables than TDM. 

Two varieties of FDM are commonly used. The first, MHz FDM, places each TES in series with a resonator with a unique resonance frequency of O(1) MHz. Each TES is independently biased, through the same bias line, with an alternating-voltage at the frequency of its associated resonator and the TES signal is amplified by a DC-SQUID array before being further amplified and demodulated by room temperature electronics. The independent TES bias allows for independent optimization of each TES operation point, accommodating large variation in TES parameters. MHz FDM systems have been fielded with multiplexing factors as high as 68 in CMB experiments \cite{bender_-sky_2020}, but the increased TES bandwidth required for \comm{a CUPID-1T scale detector} will necessitate the use of a lower multiplexing factor. Additionally, cryogenic signals are slow and do not need digitization at a rate beyond a few kHz; a modest multiplexing factor of 10 would produce a manageable number of wires. The dominant source of thermal loading on the detector cryogenic stage is from the wires required to bias and readout each multiplexed set of TESs, allowing for very low additional thermal loading. GHz FDM, usually referred to as microwave multiplexing or $\mu$MUX, provides a DC voltage bias to sets of TESs and couples their current to superconducting GHz resonators using RF-SQUIDs \cite{cukierman_microwave_2020}. A cryogenic amplification stage and further room temperature amplification and demodulation are used to retrieve the TES signal. The lack of independent TES biases can place a constraint on the uniformity of TESs that are biased with the same line. $\mu$MUX systems promise multiplexing factors of O(1,000), but have not yet been fielded on that scale. $\mu$MUX also has significant thermal loading on the coldest cryogenic stage from dissipation in the microwave resonators and heat leak through the coaxial cables and DC cables. 

However, while FDM readout may be promising, there are specific challenges that must be addressed by the community as they pertain to multi-channel experiments such as CUPID-1T:
\begin{itemize}
    \item The experiments are constructed within a dilution refrigerator with a working temperature of $\sim$0.01\,K. Multiplexing has not yet been demonstrated at this scale in the temperature range necessary for successful bolometric operation. 
    \item The radioactivity of uranium and thorium that is naturally occurring in materials used for multiplexing readouts is a background in rare-event searches and will negatively impact qubit coherence. For example, any cables that are fed into the experiment chamber must meet stringent radiopurity requirements. 
    \item Cables must also be appropriately shielded from magnetic flux. Any stray currents or fields (including Earth’s magnetic field) must be addressed when considering the meters or more of cables required for low-noise readouts. 
\end{itemize}
A demonstration of these technical solutions is necessary for the advancement of infrastructure for cryogenic readout of TES arrays in CUPID-1T and will simultaneously have high impact on the technological advancement in QIS, as described further in section \ref{sec:BroaderImpacts_QuantumComputing}. The R\&D program towards CUPID-1T foresees testing to address these challenges. MHz FDM on the order of 10 multiplexed TESs is being developed for applications to the DEMETER demonstrator project (see Section \ref{sec:Synergies_DEMETER_MUX}) with the expectation that the technology can be scaled to 10--100 thousand channels with CUPID-1T. 

        
\subsection{Light detectors with signal amplification based on the Neganov-Trofimov-Luke effect} \label{sec:NeganovLuke}
        The light detectors (LD) for \cupid and for a larger, ton-scale detector like \cupidt will play a critical role in achieving the background levels required to meet the science goals of these experiments. 

\begin{figure}
\begin{center}
\includegraphics[height=2.25in]{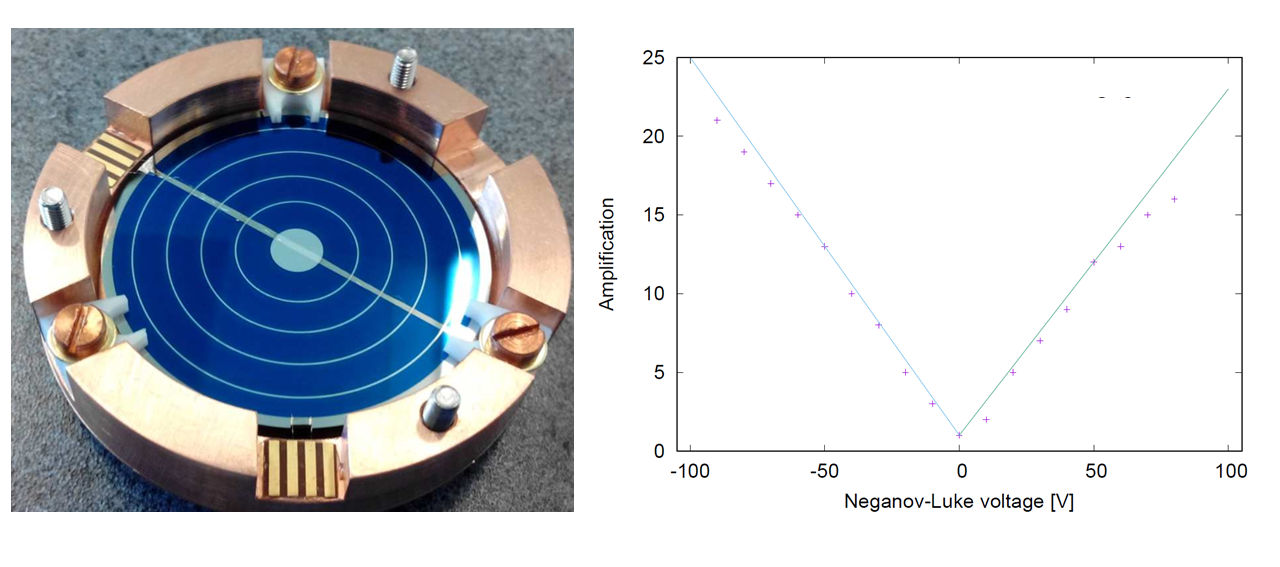}
\caption{\label{fig:Neganov-Luke}  Left: Photograph of an NTL light detector consisting of a Ge wafer with concentric Al electrodes and NTD readout. Right: The enhancement of the thermal signal as a function of the voltage across two adjacent rings is reported. Amplifications up to 20 can be achieved with voltages of $\sim$100~V.}
\end{center}
\end{figure}

The performance of a semiconductor-based bolometric light detector can be enhanced by exploiting the Neganov-Trofimov-Luke (NTL) effect \cite{Neganov:1985, Luke:1988}, namely, the amplification of the particle-induced thermal signal as a result of the collection of the charge carriers by an applied voltage.  NTL-assisted detectors are typically fabricated through the deposition of electrode(s) on the absorber surface(s) (Fig.~\ref{fig:Neganov-Luke}, left) --- using well established evaporation, sputtering, and ion implantation techniques --- in order to apply a voltage bias at low temperatures (e.g., see \cite{Pirro:2017, Novati:2019} and references therein). The extra heat induced by drifting electrons and holes in the static electric field scales linearly with the voltage applied to the device, up to a certain bias at which a leakage current occurs (Fig.~\ref{fig:Neganov-Luke}, right). The signal-to-noise ratio has a similar voltage dependence, but the maximum is usually reached at a lower voltage than that providing the highest signal amplitude. Alternative designs of NTL detectors have been proposed \cite{ Mirabolfathi:2009, Chapellier:2015, Defay:2016, Mast:2018, Defay:2019} with the goal of reducing the parasitic currents flowing in the semiconductor across the electrodes. 

Bolometric detectors with NTL-amplification have been developed for three decades, primarily for dark matter search experiments \cite{Spooner:1991, Shutt:1992, Brink:2006, Eitel:2006, Broniatowski:2008, Broniatowski:2009, Agnese:2013, Mirabolfathi:2014, Hehn:2016, Agnese:2016, Arnaud:2016, Agnese:2017, Pirro:2017, Armengaud:2017, Armengaud:2018, Arnaud:2018, Agnese:2018, Agnese:2018a, Romani:2018, Ponce:2020, Arnaud:2020, Amaral:2020}. These devices are also of great interest in experiments for the detection of coherent elastic neutrino-nucleus scattering \cite{Billard:2018, Neog:2020, Beaulieu:2021, Salagnac:2021}. Moreover, several projects involving the use of NTL-LDs for rare event search experiments have recently been developed (see the recent review \cite{Poda:2021} and references therein). 

The use of NTL-LDs for bolometric double-beta decay searches is of particular interest,  due to the possibility of achieving particle identification with non-scintillating (or poorly-scintillating) bolometers through the detection of Cherenkov radiation induced by particle interaction in the crystal.  This technique was proposed in \cite{TabarellideFatis:2010} and demonstrated in \cite{Willers2015a, Pattavina2016, Gironi:2016, Artusa2017, Berge2018}. NTL-LD prototypes with detection surfaces of about 15~cm$^{2}$ of germanium (and up to 4~cm$^2$ of silicon) have been developed and tested in both above-ground and underground facilities (see \cite{Poda:2021, Novati:2019}. Using the NTL amplification mechanism, typical NTL-LD devices achieve signal amplitudes on the order of ten(s) of $\mu$V/keV and baseline resolutions of approximately $\sim$5-20~eV FWHM, with a grid bias of up to 70~V (in germanium; 300~V in silicon).  This corresponds to a factor of 10 (50) improvement in the signal-to-noise ratio with respect to the standard germanium (silicon) based LDs. 

The importance of NTL-LDs for CUPID-1T is two-fold:	

\begin{enumerate}
\item Given the low light yield of LMO scintillators and the absence of a reflective material around the crystals in the CUPID detector structure, NTL-LDs can improve alpha-particle rejection efficiency, relaxing the demands on the detector performance at 0~V electrode bias (which corresponds to the standard LDs). 	
\item In comparison with massive LMO bolometers, the enhanced signal-to-noise ratio and faster response time of NTL-LDs makes these devices valuable for the rejection of random coincidences (which are mainly two-neutrino double-beta decay events) \cite{Chernyak2012,Chernyak2014,Chernyak2016}, which represent one of the most significant contributions to the CUPID background budget.
\end{enumerate}

The technology of NTL-LDs is mature enough to be considered for large-scale bolometric experiments like CUPID and \cupidt. However, the electrode pattern for the detector module design needs to be adapted to ease the NTL-LD production and assembly, as well as to improve charge collection. The reliability and reproducibility of the performance of NTL-LDs also needs to be demonstrated with tens of devices operated for over a year of continuous measurements.  The ability to maintain a voltage a bias high enough to achieve at least a factor of 10 in amplification of the signal-to-noise ratio on this time scale is of particular importance.  This reproducibility is critical for the validation of the NTL-LD technology and for a possible optimization of the wiring in the cryostat (i.e., applying bias on the NTL-LDs electrodes in parallel, thus reducing the number of additional wires to be used). These tasks are subjects of ongoing activities within the BINGO and CROSS projects (sections \ref{sec:ActiveGammaVeto} and \ref{sec:CoatingsForPSD}, respectively).

It is important to stress that the NTL technique can be applied to LDs with any type of phonon sensors. Even when the sensor is much more sensitive than an NTD (as in the case of TESs \cite{Willers:2015}, for example), an additional amplification of the signal could lead to more effective pulse-discrimination and improvements in pile-up rejection.

\subsection{Near-Term Demonstrators}
In addition to the results of the many projects and demonstrators discussed above, multiple ongoing R\&D projects are underway in self-contained experiments. A selection of these projects
 and their work toward applying the necessary technology to meet CUPID-1T requirements at CUPID-1T scales are described here. 

    \subsubsection{Bi-Isotope \ndbd Next Generation Observatory (BINGO)} \label{sec:ActiveGammaVeto}
        \paragraph{Active \ga Veto}
                External \ga's --- originating either from outside of the cryostat or from within the cryostat shields --- may pose a problem even for candidates with a high transition energy when extremely low background levels are required. The most challenging components are the 2615 keV line of $^{208}$Tl, but also some low-intensity characteristic \ga’s from $^{214}$Bi (belonging to the $^{238}$U chain and related to $^{226}$Ra and radon progeny) above 2615 keV. There is in particular a line at 3054 keV, but with only 0.021\% branching ratio. This background contribution can be effectively mitigated with an internal veto, located in the mK region immediately around the bolometers and with no inert material interposed.

This possibility is being explored at the R\&D level by the BINGO project~\cite{Nones:2021b}, which will use TeO$_2$ and Li$_2$MoO$_4$ crystals for a mid-scale technology demonstrator. The internal shield proposed in BINGO is an almost hermetic arrangement of scintillating bars, forming a barrel all around the arrays of crystals and placed near the bolometers. Each bar of the shield will be a single crystal based on ZnWO$_4$ scintillator.  Alternative crystal bars of BGO are also under consideration. For the intial design,  ZnWO$_4$ scintillator was chosen because of several attractive features, including high density (7.8 g/cm$^{3}$); high average $Z$ ($\sim$~51); the possibility of growing large crystals; high scintillation yield, on the order of at least 9300 photons/MeV at room temperatures~\cite{Holl:1988a} and even higher at low temperatures; long light attenuation length ($\sim$~20--30 cm) at the peak emission wavelength (480 nm); and an impressive achievable radiopurity ($<$ 0.17 $\mu$Bq/kg in $^{228}$Th~\cite{Belli:2019a}). 

Each bar will be read out by two Neganov-Trofimov-Luke Ge-based light detectors (see section \ref{sec:NeganovLuke}) at its extremities.  It is worth remarking that the ZnWO$_4$ crystals in the these tests will not be operated as bolometers, but as pure scintillators (with a major simplification of the assembly). Only light detection will be performed bolometrically, as this choice is quite convenient in a mK environment.

        \paragraph{Background Reduction Through Geometry Optimization}\label{sec:FullActiveAssembly}

Event coincidences between different detector modules is a powerful technique for background rejection. Double beta decay events provide mostly single-site signals. Simultaneous signals in more than one module can be attributed to $\gamma$, muon, and neutron backgrounds. Surface radioactivity can also be rejected with this approach, as charged particles emitted at a surface of a module can be stopped in a nearby module if there is no interposed inert material. For this reason, detector holders are designed to maximize the direct visibility between different modules, to provide as open a structure as possible. 

\begin{figure}
\begin{center}
\includegraphics[height=2.25in]{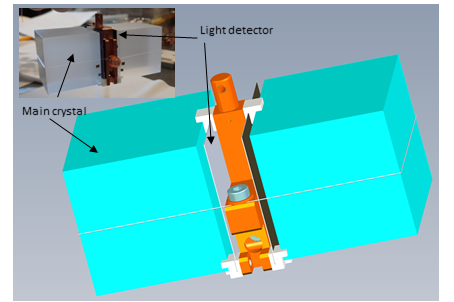}
\caption{\label{fig:BINGO-assembly} Scheme of an innovative detector construction --- adopted in BINGO --- aiming at a substantial reduction of the inert material facing the main crystal. In the top-left inset, a photograph of an actual assembly comprising 45$\times$45$\times$45~mm Li$_2$MoO$_4$ crystals and 45$\times$45~mm Ge light detectors is shown.}
\end{center}
\end{figure}

In order to dramatically reduce the amount of inert surface directly facing the crystals containing the double-beta decay candidates, the BINGO project has proposed a ground-breaking approach in the detector construction \cite{Nones:2021b}.  The primary idea is to reject most of the surface radioactivity (especially that of $\beta$ origin, since the $\alpha$’s are rejected by light collection) by relying on the geometry of the assembly and exploiting the presence of the light detectors. The basic concept consists of holding a cubic Li$_2$MoO$_4$ crystal from the side with PTFE elements that connect it to a copper bar (which acts as a heat sink for the detector), in such a way that they also hold a vertically-placed light detector with the same cross-section as the main crystal, completely shielding the copper structure. Nylon thin wires can be used to improve the crystal stability. (See Fig.~\ref{fig:BINGO-assembly}.) Events involving energy deposited into both a crystal and a light detector will be rejected if their amplitude does not comply with the expected heat-light ratio.

Mechanical tests of this structure and low-temperature bolometric measurements have been performed at IJCLab in Orsay, France, with encouraging results. Special PTFE pieces were designed (three for each crystal), to connect the crystal to the copper heat sink and simultaneously clamp the light detector, interposed between the crystals and the copper bar. The crystal was a Li$_2$MoO$_4$ element with the same dimensions as a CUPID detector (45$\times$45$\times$ 45~mm), and the light detector was a 45$\times$ 45~mm square Ge wafer. A single nylon wire with a tension of 4~kg and a diameter of 0.4~mm pressed the crystal against the copper heat sink through the PTFE elements, simultaneously clamping the Ge wafer. The energy resolution achieved on the Li$_2$MoO$_4$ element was the same as that obtained with conventional detector mounting.

The read-out for the main crystal was accomplished by gluing the NTD-Ge thermistor onto the crystal and near the edge of the side facing the light detector, so that the bonding wires reached pads placed on the copper element just by passing very close to the light-detector edge with a loop. A similar arrangement is foreseen for the heater to be used for temperature stabilization of the crystal response. No reflector will be used around the luminescent crystals and there will be only one light detector for each crystal. The negative effect that this arrangement has on the light collection calls for a substantial improvement of the light-detector sensitivity, which can be obtained either by the Neganov-Trofimov-Luke effect (as proposed in BINGO) or by using TES-based light detectors.   

This geometry can be repeated in order to form arrays. A long copper bar can support stacked detectors on opposite sides, forming a tower with two detectors per floor. These towers can be arranged in a matrix and mounted very close each other, so that each core crystal is fully surrounded (but the small PTFE and nylon details) by active elements. The reduction of the passive material surface is of the order of a factor~of~100 with respect to current construction schemes. It is interesting to observe that in this arrangement a volume is formed between the two light detectors and around the copper bar where one can put ``dirty surface elements'', since objects placed here do not directly face the Li$_2$MoO$_4$ crystals. This volume can be exploited to place readout wires and copper tubes to insert calibration sources. While these elements must be radiopure, stringent (and very difficult) control of their surface radioactivity is not required. 

This construction principle will be tested in a mid-scale demonstrator --- dubbed MINI-BINGO --- with 24 crystals in the Modane Underground Laboratory.

    \subsubsection{Cryogenic Rare-event Observatory with Surface Sensitivity (CROSS)}\label{sec:CoatingsForPSD}
        
\begin{figure}
\begin{center}
\includegraphics[height=3in]{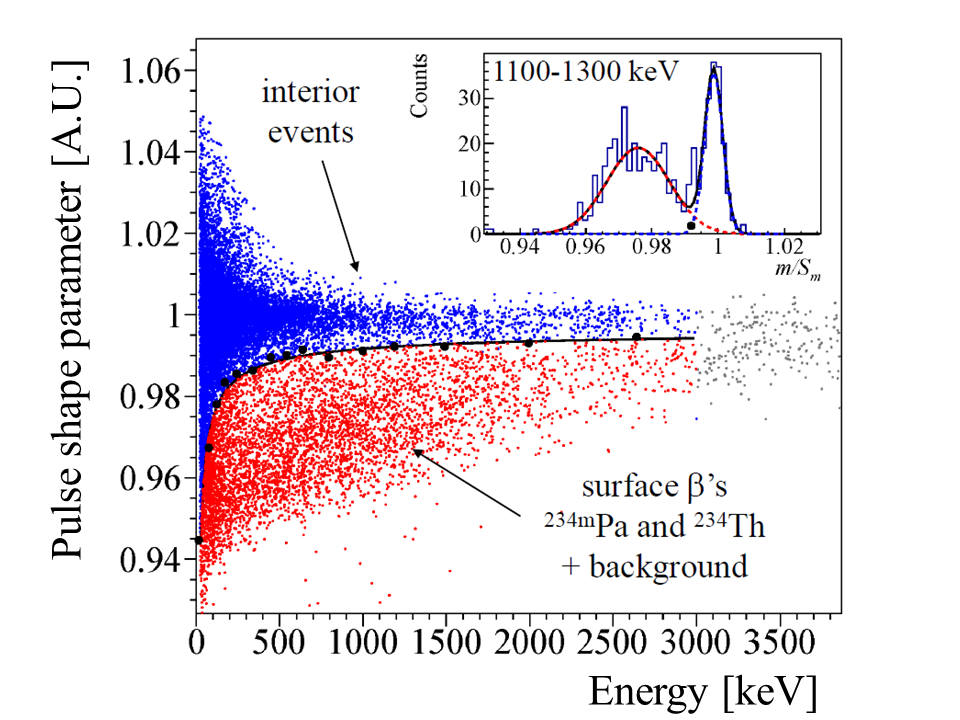}
\caption{\label{fig:beta-separation} Surface-event identification obtained by a Li$_2$MoO$_4$ detector with an Al–Pd coating exposed to a radioactive source~\cite{Bandac:2021a}. In a plot of a specially-constructed pulse-shape parameter vs. energy, the surface events (in red) produced by the source emitting $\beta$'s from the nuclei $^{234m}$Pa and $^{234}$Th lay below a black curve defining a 3$\sigma$ acceptance region for the interior events (in blue). An example of interior/surface event separation using a distribution of the pulse-shape parameter in a given energy interval is provided in the inset.}
\end{center}
\end{figure}

To achieve the ambitious background goals of CUPID-1T, it is mandatory to reject $\alpha$ or $\beta$ events induced by radioactive impurities located either close to the surface of the crystal itself or to that of the surrounding structure. Surface $\alpha$’s are rejected by particle identification based on a light-yield cut. We remark however that the absorption of energetic $\beta$ particles associated to the decay of $^{214}$Bi and $^{208}$Tl, belonging to the natural radioactive chains of $^{238}$U and $^{232}$Th respectively, can contribute to the background. The rejection of these surface $\beta$’s may require dedicated techniques. One possibility is tagging surface events by pulse-shape discrimination, assisted by a proper film coating of the crystal faces. This method is investigated by the CROSS experiment~\cite{Bandac:2020a}. Efficient rejection of $\beta$ events (higher than 90\% up to $\sim$~2~MeV $\beta$ energy) was demonstrated in small prototypes based on Li$_2$MoO$_4$ and TeO$_2$ crystals (with volumes of 4~cm$^3$)~\cite{Bandac:2021a}, as shown in Fig.~\ref{fig:beta-separation}.

The concept of the discrimination capability is based on the properties of non-equilibrium phonons emerging from a hot spot generated by a particle or nuclear event in a non-metallic material. These phonons have initially (after a few $\mu$s) ``high'' energies ($> 30$~K) in a cloud of $\sim 1$~mm size around the event location. They will reach the film-coated surface when the event is located within $\sim 1$~mm from the surface. (We stress that this is the typical range of a $\sim$~MeV  $\beta$~particle emitted by surface radioactive impurities.) When the energy deposition occurs at a longer distance from the surface (bulk events), the particle-generated phonons reach the film with a significantly lower average energy, due to the quasi-diffusive mode of phonon propagation, which implies spontaneous fast phonon decay when the phonon energy is high. Since the metallic film plays an important role in the thermalization time of the absorbed phonons, signals from events occurring close to the film will present a modified time evolution with respect to those taking place in the bulk. If the film is a normal metal, the energy deposited by the phonon in it is quickly thermalized in the electron system, so that phonons of much lower energies are re-injected in the crystal from the film. Signals from events occurring close to the film will present, therefore, a shorter rise time if the detection is mediated by thermal phonons as in the case of an NTD-Ge sensor. If the film is a superconductor, the out-of-equibirum phonons break Cooper pairs and form quasi-particles, which present a long life time (up to several ms) before recombining back to phonons, so some energy is trapped in the film. This mechanism competes with the former previously described. 

In real experiments with prototypes, faster pulses (with a shorter rise-time) from surface events were observed with normal metal films (Pd, thickness 10 nm) and low-gap superconductive films (Al-Pd bi-layers, 100~nm/10~nm thick respectively, critical temperature 0.65 K)~\cite{Bandac:2021a} (see Fig.~\ref{fig:beta-separation}). With Al films (critical temperature 1.2~K), results are less convincing. First of all, only surface \al's are clearly rejected~\cite{Bandac:2020a}, so this method is not interesting as it adds nothing to light-yield cuts. Secondly, the effect of a larger gap and a more effective energy trapping is visible, as the surface events produce slower pulses (with a longer risetime) for some types of phonon sensors, mainly sensitive to out-of-equilibrium phonons~\cite{Nones:2012a,Bandac:2020a}. Therefore, the current protocol to achieve surface sensitivity --- still to be transferred to large crystals --- is based on a surface coating with Al-Pd film in the form of a grid~\cite{Bandac:2021a}. This configuration was also tested and shown to work. The advantages of the grid consist in a further reduction of the heat capacity of the coating, the possibility to extract scintillation light and the availability of geometrical parameters (pitch and widths of the grid lines) to possibly tune the discrimination parameters.

    \subsubsection{Demonstrator Experiment for Multiplexed Event Topology and Energy Reconstruction (DEMETER)} \label{sec:Synergies_DEMETER_MUX}

A ``background-free” operation of the CUPID-1T detector is key for reaching the ultimate sensitivity.
One mechanism to accomplish this could be the topological reconstruction of events at the single crystal level, i.e. directional or spatial discrimination of single-beta and double-beta events. Reconstruction using a combination of light propagation and phonon wave detection would result in total event discrimination, as a \ndbd\  would have a light: heat signature that is unique to the decay. Unfortunately, topological reconstruction inside solid bolometric crystals has not yet been achieved. CUORE has modeled the thermal response from CUORE phonon signals, but does not yet have the ability to reconstruct events based on single-NTD sensors; however some modifications could make this possible. ``Phonon imaging” has been illustrated in TeO$_2$ which is a candidate \ndbd\ bolometer material \cite{hurley_phonon_1986}, using additional instrumentation and comparisons with simulated phonon wave propogation. Photon reconstruction is common practice in gaseous or liquid \ndbd\  detectors, including the use of machine learning techniques for light response calibration in liquid xenon \cite{hansen_radon_2019}. A successful demonstration of this combined technology would  dramatically improve capabilities to search for rare events like \ndbd;
topological discrimination of \ga and \be in \teo could mitigate the need for enriched detectors for experiments like \cupid and \cupidt.  
It would also contribute to searches for dark matter which rely on similar technology. The sensitivity of such a detector to the kinematics of \ndbd\  process (angular/energy distributions of the detected electrons) also opens the door to searches for the mechanism of \ndbd\  should it be detected, a capability that will dramatically strengthen future searches. 

\begin{figure}
    \centering
    \includegraphics[width=0.5\textwidth]{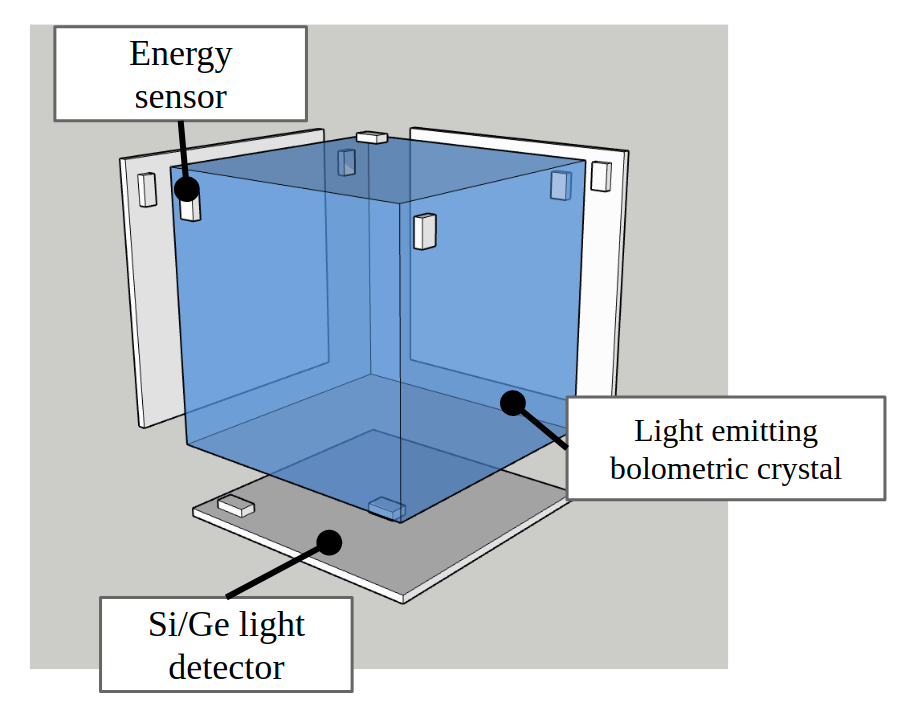}
    \includegraphics[width=0.4\textwidth]{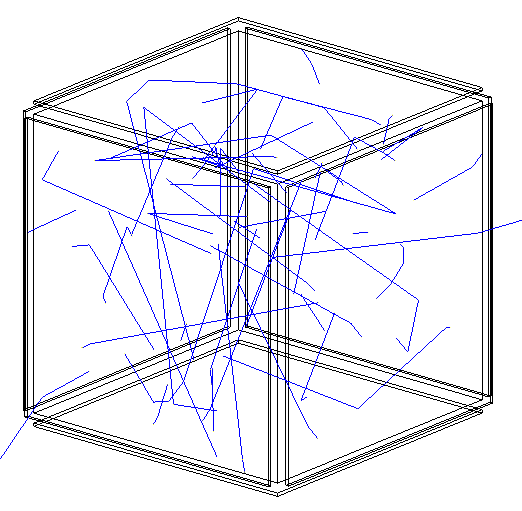}
    \caption{\emph{Left:} Cutaway of proposed DEMETER setup (infrastructure not shown, exploded view). A light emitting bolometric crystal will be instrumented on all sides with energy sensors to collect phonon readout. Each side will also face a secondary light detecting bolometer (three of six sides shown). \emph{Right:} Simulated geometry of DEMETER in Geant4, demonstrating ongoing development of light propogation simulations. }
    \label{fig:RD_MUX_DemeterSketch}
\end{figure}

DEMETER (Demonstrator Experiment with Multiplexed Event Topology and Energy Reconstruction) is currently under development at the 24-channel scale to produce scalable multiplexing technology required for the ten thousand channel order of next-generation experiments and technical solutions required for topological reconstruction of events in cryogenic crystalline bolometric experiments. Event propagation simulations are in development using existing Monte Carlo frameworks (Geant4 \cite{agostinelli_geant4_2003} and RAT-PAC \cite{seibert_stan_rat-pac_nodate}). For phonon wave propagation, the cubic bolometric crystals must be instrumented on each face with energy sensors. To collect light signal in the final demonstrator, each crystal will be instrumented with secondary light-detecting bolometers (see Figure \ref{fig:RD_MUX_DemeterSketch}). 

UC Berkeley and Lawrence Berkeley National Lab (Nuclear \& Physics divisions) head these efforts. Detectors will be constructed in collaboration with Marvell Nanofabrication Laboratory at CITRIS at UC Berkeley and Argonne National Laboratory. This design is intended to be modular --- indeed, this demonstrator will also provide an excellent test stand for other light-detecting technologies in the case that a better design needs to be characterized \emph{in situ}.
Plans are already in place to implement tests with other light-emitting crystals, namely with Cherenkov light from \teo. 
In the current design, each crystal face and secondary light-collecting bolometer will be instrumented with TES readouts. As described above, TES technology has been demonstrated to meet threshold and timing resolution requirements for CUPID \cite{singh_vivek_development_2020} using the dilution refrigerator at UC Berkeley, where many of the preliminary tests for this project will take place. Alternatively, superconducting nanowire single photon detectors have also demonstrated excellent time resolution and their cryogenic and radiopure operation may apply well to the requirements of DEMETER \cite{polakovic_unconventional_2020}. This brings the total number of energy sensors to twelve per crystal, which will be read out by a limited number of SQUIDs, driving the application of multiplexing. SQUID array readout is currently under development, in collaboration with LNBL Cosmic Microwave Background experimentalists and in parallel with evaluation of components for the final underground design. Further requirements for multiplexed readout on CUPID-1T are discussed in section \ref{sec:Multiplexing}.

    \subsubsection{Low-Impedance TES \& Synergies with Ricochet} \label{sec:SynergiesWithRicochet}


TES sensors have also been employed on macro bolometers in rare event search experiments for several decades  in CRESST\cite{Bravin:1999} and in CDMS  \cite{Barnes:1993} and are considered as an option for the readout of the \lmo absorber in \cupidt. Traditionally for these detectors the TES sensor is deposited directly onto the target allowing to obtain sensitivity to the athermal phonon signal with both excellent timing and energy resolutions. Recent devices have demonstrated energy resolutions down to 2.7 eV with 20~$\mathrm{\mu}$s rise-times on a 1~g Si absorber \cite{Ren:2021} and 4.6~eV resolution on a 24~g CaWO$_4$ target \cite{Abdelhameed:2019}. 
However, the prospect of combining the stringent radio-purity control requirements for \cupidt with an elaborate sensor litography/deposition program on each one of the $\mathcal{O}(10^4)$ detectors is daunting. 
The TES-based detector design discussed in the following hence avoids the deposition of the TES sensor on the target and instead couples the target to the sensor through a simple Au deposition together with Au wirebonds to ensure thermal connections. This technology has been developed independently for the Ricochet Coherent Neutrino Nucleus Scattering experiment (\CEvNS) \cite{Ricochet:2021a}, and is very similar in its concept to the AMoRE MMC sensor based readout \cite{Alenkov:2019jis,Kim:2017} and a recently presented TES based architecture, called the remoTES \cite{Angloher:2021}.
Although not fabricating the TES sensor directly on the absorber does decrease both the sensitivity to athermal phonons and the ultimate theoretical energy resolution of the device, the appeal of the technology is its easy adaptability to different absorbers, its scalability with $\mathcal{O}(10^3)$ sensors being produced on a single standard 6" silicon wafer, and its compatibility with multiplexed SQUID readout solutions presently being used for high pixel count CMB observatories. Most importantly, optimizing the design allows for the prospect of achieving all of these features while still achieving fast signal rise-times compatible with the background requirements for \cupidt.
Due to this broad range of features also appealing for the larger community, \cupidt is expected to profit from ongoing technology developments and in particular from a large overlapping set of requirements for next generation \CEvNS experiments. Experiments like NUCLEUS\cite{Angloher:2019} and Ricochet\cite{Ricochet:2021a} are poised to take first data with $\mathcal{O}(10)$ detectors and provide first precision measurements of \CEvNS from reactor neutrinos in the coming two years. Both collaborations have expressed a compelling strategy to extend their payload of g-scale or 10 g-scale detectors to a total mass of 1~kg (few tens of 10~kg), which will require $\mathcal{O}(10^3)$ detectors. We hence expect significant ongoing efforts to further develop and improve key technology aspects like multiplexed readout and optimal sensor coupling for fast signal rise- and decay-times. For \CEvNS experiments the fast timing avoids dead-time for above ground detector operation, and a signal observation requires thresholds in the tens of eV. For the \ndbd search in CUPID, the timing allows to reject pile-up backgrounds and an improvement in energy resolution leads to a smaller ROI.

\begin{figure}
\begin{center}
\includegraphics[height=2.25in]{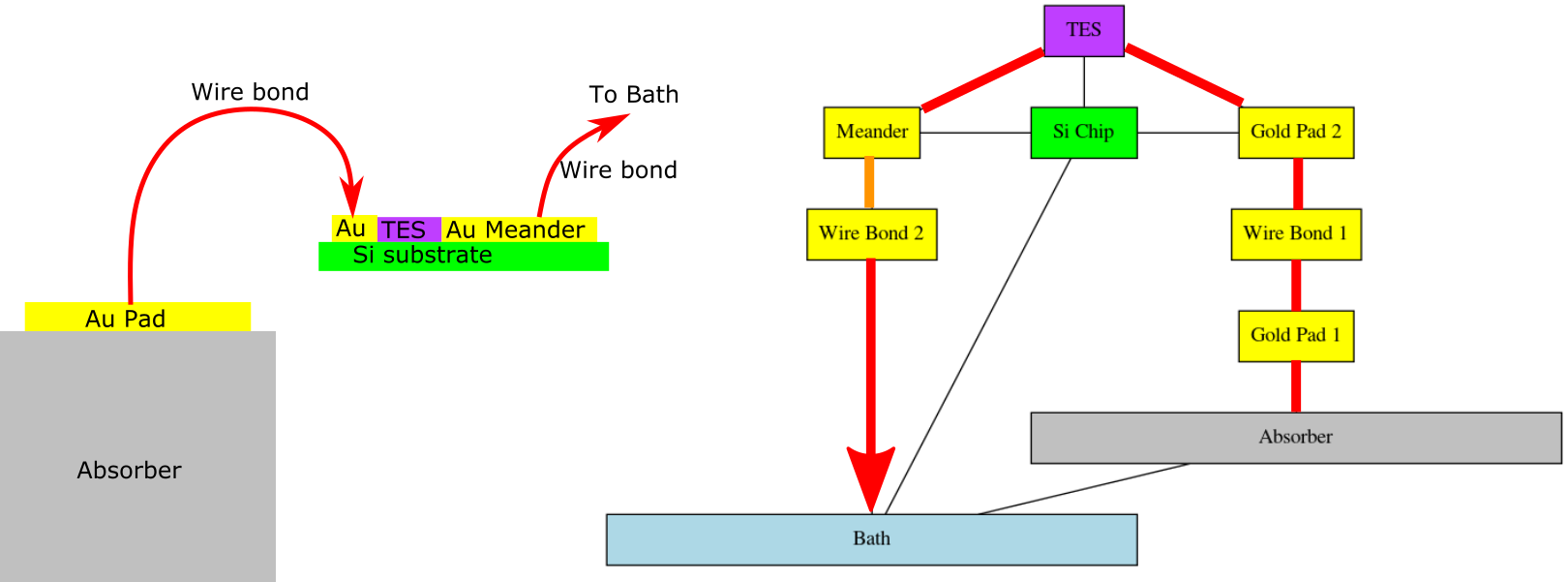}
\caption{\label{fig:ThermalTESDiagram}  Left: Schematic representation of the heat flow through our detector conocept separating the target from the TES sensor chip. Right: Thermal model block diagram of the absorber and TES architecture. Each component is modeled with a heat capacity indicated by a block, and a thermal coupling (line) is modeled between each component.  The heat flows primarily through the Au path highlighted in red. The energy resolution and decay time constant of of the TES is largely set by the `Meander' portion of the path shown in orange, which can be tuned for different targets.  }
\end{center}
\end{figure}

In the following we will present the status and expected resolution of a CUPID calorimeter with the thermal modeling and design of the TES architecture developed for the Ricochet experiment. The architecture is based on early modeling work done in~\cite{Figueroa:2006} and applied to small rare-event search detectors in~\cite{Pyle:2015pya}. Fig.~\ref{fig:ThermalTESDiagram} shows a schematic of the idea behind the architecture on the left and thermal model block diagram on the right. The absorber is isolated from the thermal bath by low thermal conductance sapphire supports \cite{Pinckney:2021} or other dielectrics with negligible heat conductance. The separate TES chip, deposited on a Si substrate, is mounted on the housing above the target and thermally connects to the target via Au wirebonds. A gold meander deposited onto the same TES chip and connected to the bath through another Au wirebond is used to control the thermal connection from the TES to the bath. Its conductance is tuned to ensure that there is no phase separation of parts of the TES and allows us to optimize the decay-time and energy resolution. In the thermal block diagram in Fig.~\ref{fig:ThermalTESDiagram} (right) heat capacitances C are denoted as blocks and thermal conductances G as lines. The dominant thermal conductance and the path of the heat flow are highlighted in red. 
\begin{figure}
\begin{center}
\includegraphics[height=2.5in]{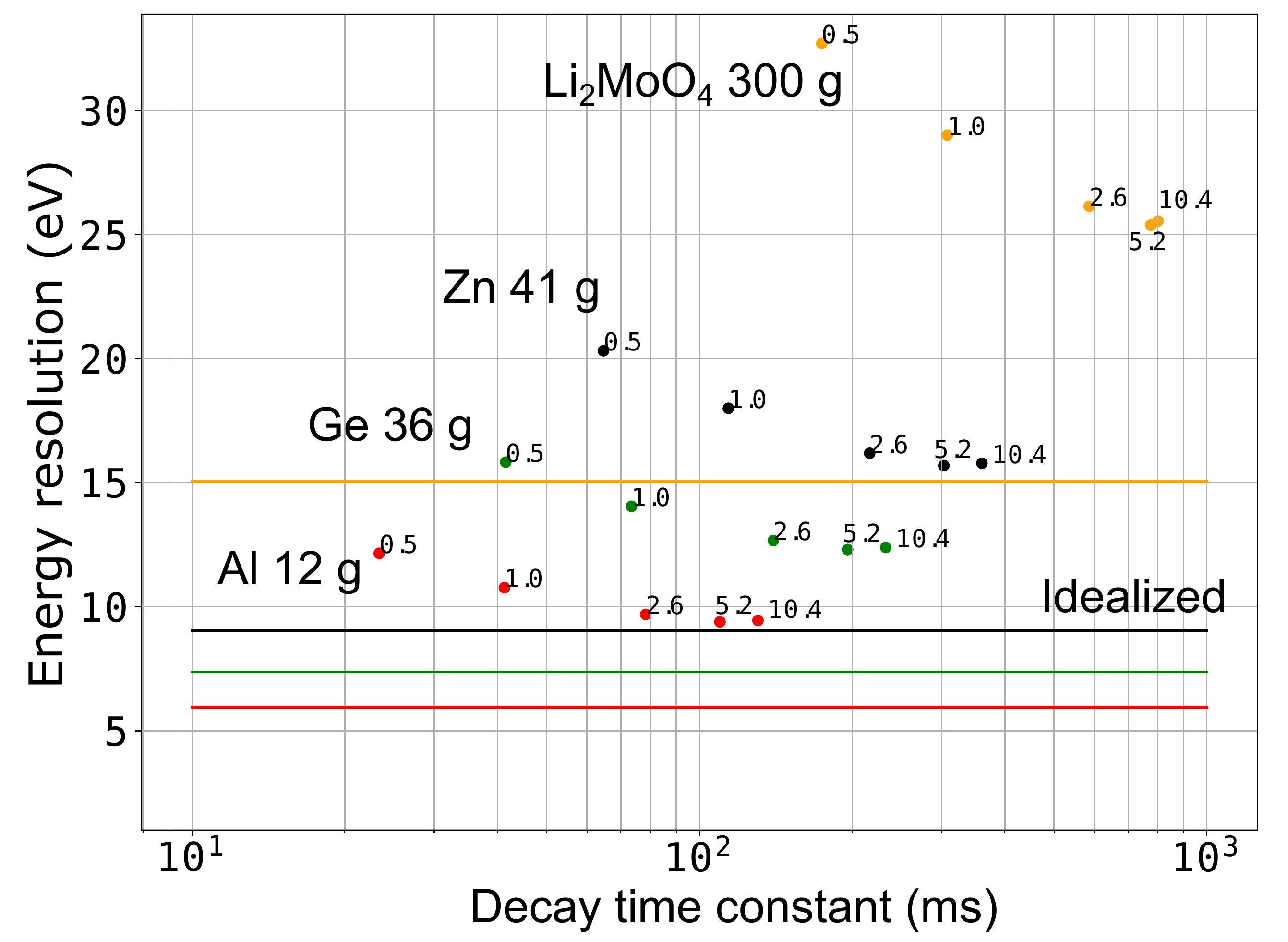}
\caption{\label{fig:TESResolution} Numerical thermal model predictions for the energy resolution of TES-based thermal chip designs for a 300~g \enrlmo target as well as for potential Ricochet targets TES detector as a function of modeled  decay time constant.  The meander length is indicated (in mm) for each of the evaluated designs.   For comparison the energy resolution of an idealized model with a single heat capacity of the absorber only is given as a solid line.}
\end{center}
\end{figure}
Values for common heat capacities and conductances have been taken from the literature \cite{Pobell:2007, Gopal:2014, Lorentz:2019} with the \lmo Debye temperature from \cite{Musikhin:2015}. Dedicated studies were performed for thin films as for example for the 300~nm thick Au meander \cite{Chen:2021} or for other components that will be dominated by boundary resistances like the sapphire balls intended as weak thermal links \cite{Pinckney:2021}. The block diagram has subsequently been translated into a set of ordinary differential equations and implemented in a non-linear solver in python to simulate the expected signal response in terms of pulse shape and energy resolution.  The model predictions have been validated against a preliminary TES chip with a critical temperature ($T_c$) of 80~mK and the model has been used to optimize a subsequent TES chip design. Further details of the thermal model have been described in \cite{Chen:2021}.
Predicted baseline energy resolutions and decay time constants for different targets for CUPID and Ricochet are shown as a function of  Au meander length in Fig.~\ref{fig:TESResolution}. A simplified idealized resolution estimate for each target is shown as a horizontal line. This idealized model treats the system as a single heat capacity $C_\mathrm{tot}$ set to the sum of the heat capacity of all elements, thus represents a ideal scenario where all the internal conductances are essentially infinite. All models assume a TES critical temperature of $T_c=40$~mK and a TES responsivity $\alpha = 100$. The idealized resolution is calculated following \cite{Figueroa:2006} as
\begin{equation}
\sigma =\sqrt{\frac{4\cdot k_b\cdot T_c^2\cdot C_\mathrm{tot}}{\alpha}}.
\end{equation}

\begin{figure}
\begin{center}
\begin{tabular}{c c}
\includegraphics[width=0.95\textwidth]{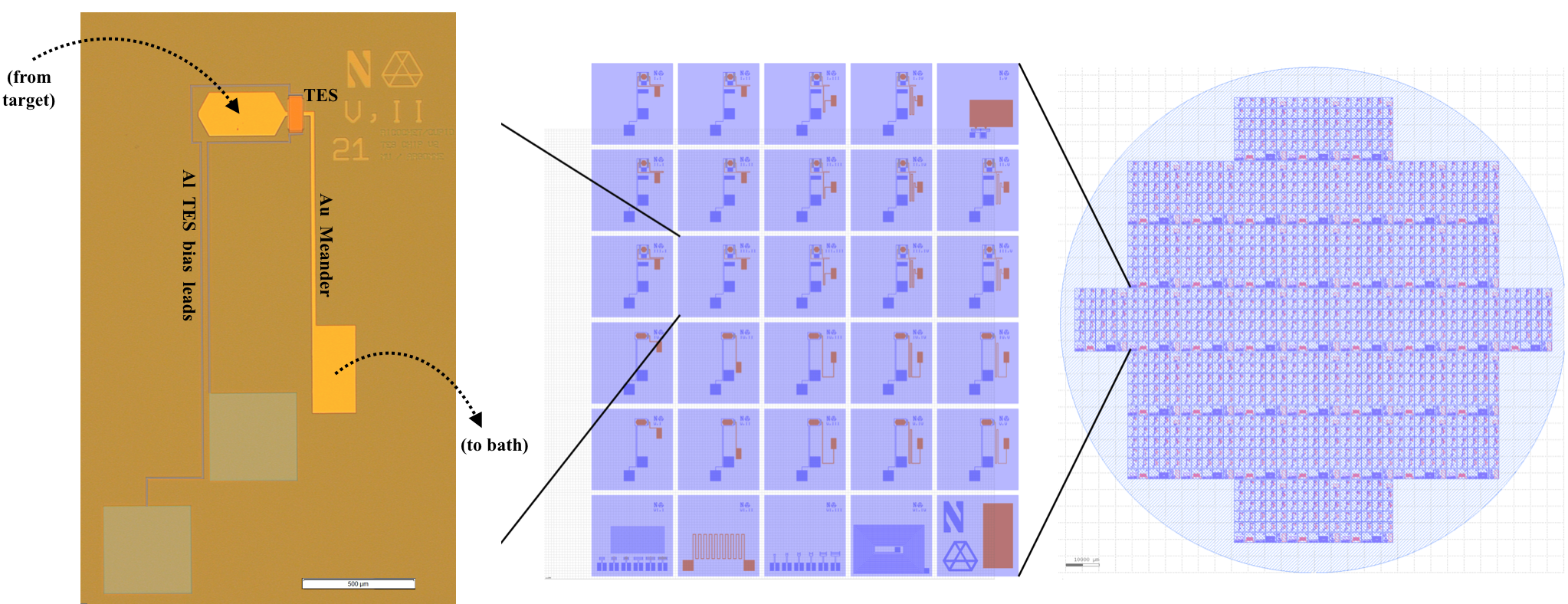}
\end{tabular}
\caption{\label{fig:TESonWafer}  Left: A photograph of a single TES chip developed by NU and ANL for the Ricochet experiment. Heat flows into an AlMn TES from the Au pad in the upper left, and out of the TES into the bath through the Au pad in the lower right.  Al TES bias leads have minimal thermal effect. Middle: A layout rending of one `block' of 25 diverse TES chips and 5 quality control samples. Right: Layout rendering of the full mask, containing 1,075 TES chips on a single 6 inch wafer. }
\end{center}
\end{figure}

The predicted baseline energy resolution of down to $\sim$25~eV (15~eV idealized) for 300 g \lmo detectors is better than any presently demonstrated detector performance with NTD based detectors \cite{Armengaud:2019loe}. Assuming we can achieve the predicted performance to within a factor of a few, this detector design is expected to be competitive with respect to NTD based detectors in terms of energy resolution while allowing for the tuning of the rise-time down to the thermalization time of phonons in \lmo.

A prototype detector with a TES chip developed by Northwestern University and Argonne National Laboratory has been operated in december of 2021 with a 1~g Si absorber as target for the Ricochet program. Pulse rise-time constants of $\mathcal{O}(100\,\mathrm{\mu s})$ have been observed and we are presently performing further studies optimizing the operation conditions to fully evaluate the detector energy resolution and to perform a quantitative comparison with our model predictions. Further cryogenic tests with different targets including \lmo and a detailed discussion of the results are expected later in 2022. 

The active R\&D program will re-evaluate the model assumptions and further study the detector response. We will assess event topologies for events in the absorber (1), the Au film (2) and the Si substrate of the TES chip (3). Pulse shape differences will be investigated in terms of bulk/surface event identification and energy mis-reconstruction. Effects like a partial sensitivity to athermal phonons as suggested in the AMoRE detector design discussion \cite{Kim:2017} and hence increased position dependence will be evaluated and if needed mitigated through analysis techniques or adjustments in the design of the detector architecture. Dual readout on opposite sides of the detector, the optimization of the surface film thickness and coverage or additional interface layers to avoid an acoustic mismatch and or improve adhesion of the film are design changes to be investigated. Similarly the film quality, typically characterized by its Residual Resistivity Ratio (RRR) between 300~K and 4~K and the geometry of this film is  going to be further optimized and investigated in terms of its phonon collection efficiency and its pulse shape modifying properties. The latter study could be further extended to additional metallic or superconducting films to further enhance surface event sensitivity as presently studied by the CROSS experiment with NTDs \cite{Zolotarova:2020suz}.



\section{Broader Scientific and Technological Impacts: Crossovers with Other Fields}


The discussion of several of the many possible high energy particle physics searches in Section \ref{sec:physics} and the discussion of detector and instrumentation R\&D in Section \ref{sec:randd} describe but a few of the overlaps between fundamental physics research in nuclear and particle physics and the shared instrumentation and techniques that have driven developments in both fields.  The fundamental nature of this research may lead, among other things, to breakthroughs in understanding neutrino properties and the identification of dark matter.  High precision studies of double beta decay will also provide important information about nuclear matrix elements, impacting nuclear theory. Measurements of and searches for cosmological remnants like the cosmic neutrino background, the hypothesized cosmic axion background, or other astrophysical relics, or the observation of supernova neutrinos will impact cosmology, nuclear astrophysics, and stellar astrophysics, as detections, exclusion limits, and precision measurements of particle properties are folded into our understanding of the cosmological Concordance Model and the physics of the early universe.  In time, a particle physics observatory sensitive to astrophysical events could even provide valuable contributions to the burgeoning field of multi-messenger astronomy.  Many of these searches, such as those for beyond \sm processes, may also pave the way for new discoveries in particle physics. 



 On a practical level, the adaptation of multiplexing techniques to the detectors in a segmented ton-scale bolometer may also find broader applications.  The continuous, simultaneous, wireless monitoring and readout of tens to thousands of nearly identical macroscopic devices could have a large impact in places where electronics wiring has an impact on the scalability of miscellaneous systems.  
 Studies of detector response may lead to a better understanding of solid state matter, semiconductors and materials research.  

These are some of the potential broader impacts of studies possible in a detector such as \cupidt.  Below, we discuss in detail a few of the more specific ways in which current R\&D overlaps with the interests of other fields.

\subsection{Crystal growth and characterization of crystals} \label{sec:BroaderImpacts_CrystalGrowth}


Studies of \LiCo{} and \moothree{} purification and \lmo{} crystal growth have been concentrated at the SIMaP Laboratory in Grenoble, France. 
 
In the French ANR-funded CLYMENE R\&D project (2017-2022, http://clymene.in2p3.fr/), the best results achieved for the purification of:

\begin{itemize}
    \item \LiCo{} powder (Fox Chemicals, 5N purity) by carbonation route is $\sim$0.1 ppm K \cite{Wu:2012}; 
    \item \moothree{} powder (Alfa Aesar, 5N5 purity) by ammoniac route is $\sim$0.55 ppm K \cite{Gileva:2017jfm}.
\end{itemize}

Both processes, developed on 25g powder batches, required a threefold purification loops cycle to reach a mass yield of 93 \% with K purification ratios of 16 and 5, respectively. The overall cycle took 5 days to perform for both processes carried out simultaneously. These numbers mean that the radiopure \LiCo{} and \moothree~ powders maximum production rate estimate is $\sim$8.9 kg/bench/worked year. These contents lead to an overall minimum 0.17 ppm K in the initial \lmo{} growth load, which should correspond to a 40K activity of $\sim$11.8 mBq/kg. These simple purification processes lead to the growth of crystals which are the W- and Zn-purest crystals known to date, as proved by chemical analysis and the absence of any signal at 58 K on T-I thermostimulated luminescence contour plots measured on \lmo~ single crystals \cite{Spassky:2015}. The purification processes are being scaled up to 140 g (\moothree) and 70 g (\LiCo) powder batches, without time duration increase of the overall purification cycle, within the framework of a complementary R\&D program called PULMO and funded by the Linksium Technology Transfer Accelerator Office of the Grenoble area in France (2020-2023). 70 g batches of 95\% 6Li-enriched \LiCo~ were successfully purified by the hereabove mentioned process, that is, a three-loop cycle of the same total duration, with an overall mass yield improved up to 97\%, a K purification ratio of 7.5 and a final K purity 4 ppb, and total irrecoverable 6\LiCo{} losses 1\% of the total mass injected in the successive purification loops.

A 230 g “natural” \lmo{} single crystal was first grown by the Czochralski method in a Pt crucible under air atmosphere in an unoptimized setup \cite{Velazquez:2017}. The crystals’ characterization demonstrated their promising properties for heat-scintillation cryogenic bolometer operation in terms of radiopurity levels (40K$\leq$47 mBq/kg, 226Ra$\leq$0.37 mBq/kg, 232Th$\leq$0.21 mBq/kg, 228Th$\leq$0.27 mBq/kg) and optical transmission ($\alpha$ABS(589 nm)$\approx$0.05 cm-1) \cite{Velazquez:2017}. Bolometer operation firmly established that \be, \ga and cosmic muon events can be discriminated, as well as \al-events arising from the ${}^6$Li(n,t)\al reaction. The subsequent \lmo{} cryogenic detectors also permitted to achieve a high-energy resolution, particularly for 2 keV FWHM baseline noise of a 13g-bolometer, the measured energy resolution in the energy range of 0.2–5 MeV is 2–7 keV FWHM. Thanks to high scintillation properties (one of the highest light yield, 0.97 keV/MeV, ever measured with \lmo{} scintillating bolometers), the coupling of a standard performance light detector (with 0.2–0.3 keV FWHM noise) to \lmo{} bolometers allowed to get the rejection of \al-induced background on the level of 10$\sigma$ \cite{Buse:2018}. The CLYMENE and PULMO R\&D is ongoing to develop crystal boules of ~0.9 kg (to cut three pieces of ~300 g) by means of combined numerical simulations and well-designed implementations of the \lmo{} crystals Czochralski growth process \cite{Stelian:2018,Stelian:2019}, which currently has an improvable crystallization mass yield of 81\%. 

Since 2018, several 300 to 900 g natural or 6Li-enriched \lmo{} crystals have been consistently produced, at pulling rates reaching ~2 mm/h, showing the reproducibility of this Czochralski pulling technology \cite{Stelian:2018,Stelian:2019}. Aboveground bolometer operation performed at CSNSM on 245 g crystals led to fast detectors ($\tau_d\sim$70 ms), with sensitivities up to 250 nV/keV at T=17 mK, good energy resolutions ($\sim$9-10 keV FWHM on alphas and neutron capture line) in a huge pile-ups rate configuration, low contamination (if any, no obvious alpha lines in the spectrum) and nominal light yields $\sim$0.5 keV/MeV tested with a Germanium-on-Sapphire optical detector and a too big NTD to optimize this measurement. Such remarkable performances of the crystals reflect their excellent quality: absence of twins and dislocation densities in the range 0.8–1.4 $\times$ 104 cm-2 for the crystals 50 mm in diameter. It was recently established \cite{Ahmine:2022,Ahmine:2022a} that \lmo{} crystals have a low mechanical hardness which decreases when temperature increases and that, besides, in optimized uniaxial compressive tests, the crystals crack without dislocation multiplication and without twin formation, at 450 and 650°C under stresses up to 10 and 7.6 MPa, respectively. So, the fact that \lmo{} exhibits an obviously fragile cracking behaviour without plastic response entails that, provided the crystals are grown in a stress field that remains below the minimum crack formation threshold, they necessarily exhibit low dislocation concentrations and no twins.

A 2 mm/h pulling rate corresponds to a mass uptake rate of 11.5 g/h, which turns out to be a relevant parameter in view of mass production. 1500 crystals of 300 g each can be extracted from 500 crystals of dimensions H=15 cm x f=5 cm. A tentative number of crystals annually produced by a single furnace is 20. So, with ten furnaces and a pulling rate of 10 g/h, 500 crystals can be produced in 2.5 years. The PULMO R\&D program aims at creating a startup company in 2023, which will be capable of producing \lmo{} crystals for the CUPID demand and for different rising markets which have been identified through a thorough business development survey carried out in 2020.

\subsection{Machine Learning \& Statistical Learning} \label{sec:BroaderImpacts_MachineLearning}
        \subsubsection{Event building / reconstruction / signal requirements:} 

        Energy reconstruction in CUORE uses a low threshold trigger from the optimum filter (OF) technique \cite{domizio-JInst-Loweringenergy-2011} which assumes a) that noise is stationary, and b) that energy deposits produce a linear response in pulse shape. CUORE has started to use principle component analysis (PCA) in recent analyses \cite{adams-ArXiv210406906NuclExSubmittNat-Highsensitivity-2021} for pulse shape discrimination in order to eliminate pileup events or non-physical pulses. Current work suggests that PCA can be used more broadly, as a replacement for OF, which will be explored further as CUPID-1T is developed. 
        
        CUPID also has strict constraints on background originating from the random combination of 2 single \dbd events (otherwise known as "pileup"). Rejection on the order of 1 ms can be achieved, but requires advanced software techniques. A convolutional neural network (CNN) is under development and application --- to both Monte Carlo simulated pulses and data from demonstrator runs at LNGS --- shows promise. 
        
        Finally, synergy with DEMETER (section \ref{sec:Synergies_DEMETER_MUX}) also anticipates topological reconstruction of events at the single crystal level based on the combination of heat and light signals. If successful, directional or spatial discrimination of single- and double-beta events would produce a background-free measurement. This project is still very preliminary, but will develop alongside deployments of dedicated demonstrators at UC Berkeley (TES-based) and Virginia Tech (NTD-based).

\subsubsection{Detector operation automation: } 
        
        CUORE data-taking currently involves many interactive steps throughout the shifting process. Careful tuning to the optimal NTD thermistor bias current is required for detector performance --- an automated process is currently in place to select a working point\cite{ALFONSO2021165451} --- but the upgrade from 988 to tens of thousands of channels will require more advanced procedures. It is evident that CUPID-1T should expect spreads in detector characteristics due to non-uniformities in detector and response electronics, and human intervention should be avoided. 
        
        Similarly, the CUORE Online/Offline Run Check (CORC) software \cite{Gladstone:2017} allows CUORE shifters to examine data in real time and flag time intervals where detector parameters have shifted dramatically, such as a seismic event. This is manageable for the 988 channels in CUORE but quickly becomes chaotic with tens of thousands of channels. Automated systems for assigning instabilities across data streams will be a requirement for the next generation of CUPID --- we will join a community very active in this area.

\subsubsection{Detector-wide effects: } 
        \paragraph{Data Reduction} 
                \cupidt will result in many datasets with tens of thousands of channels, priming it for the application of ongoing development in the field of data reduction algorithms. The DOE has recently funded nine projects \cite{DOEOfficeOfScienceInvestment:2021} that look specifically at processing, moving, and storing the enormous amounts of scientific data produced by the country's research infrastructure --- including scientific collaborations like \cupid and \cupidt. Topics of data transfer between member countries, as well as parallelization of computing resources will require advanced development. 
        
        \paragraph{Adaptive Algorithms for Noise Reduction}
                Recent studies have shown that vibrational noise can manifest as microphonic noise in the CUORE channels in a manner which degrades the performance of the optimum filter during analysis. The sources of this vibrational noise are extensive, but they include the pulse tubes used to cool the CUORE cryostat, seismic activity, and human-induced vibrations near the CUORE detector (e.g. footsteps and closing doors). 
                In practice, 
                this system is likely non-linear since the mechanical structure of the CUORE cryostat is incredibly complex. While it is possible to construct algorithms to estimate the quadratic transfer function, the non-linear problem quickly becomes intractable. In this case it may be possible to use machine learning methods such as recurrent neural networks to estimate the nonlinear correlations between the detector channels and the accelerometer devices \cite{SANGIORGIO2020110045}. This may provide insight into the effects of noise on trigger thresholds, triggering rates, and false positives, as well as potential correlations between cryogenic parameters and detector performance.

\subsubsection{Mid- / High-Multiplicities: } 
        
        \cupidt will achieve significant exposure with high granularity. Multiplicity in \cupid is defined as the number of crystals assigned to a certain event, and relies on multiple detector parameters including timing resolution, detector threshold, and distance between activated detectors. \cupidt will search for \ndbd at the M1 and M2 (or one- and two-crystal, respectively) level, but will experience events with higher multiplicity. Such events will either be used to constrain background rates or will be rejected as non-physical. Machine and statistical learning techniques may allow further analysis of these events for both constraining the \ndbd search as well as searching for processes beyond \ndbd. 
        
        CUORE is already exploring this space with track-like event reconstruction meant specifically for studies of muon trajectories\cite{yocum2022muon}, but such algorithms can also be generalized to broader searches. For example, ongoing work suggests similar techniques may be applied to the search for tri-nucleon decay (see section \ref{sec:OtherSpectra_TriNucleonDecays}), which simulations suggest interacts with several hundred crystals at once. Intelligently weighted track reconstruction, or development of an adaptive image recognition algorithm may increase sensitivity to physics beyond \ndbd. 
        
        \begin{figure}
            \centering
            \includegraphics[width=0.3\textwidth]{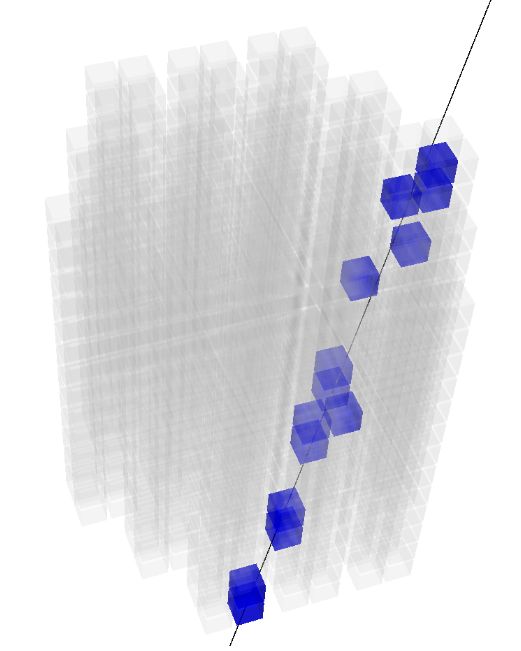} \hspace{0.12\textwidth}
            \includegraphics[width=0.3\textwidth]{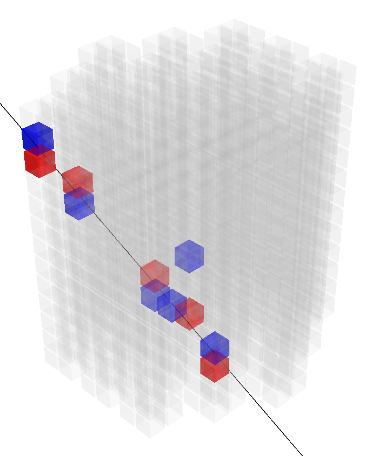}
            \caption{\textit{\textbf{Left:}} Simulated particle trajectory from toy MC created for track-reconstruction algorithm development. \textit{\textbf{Right:}} Track-like event observed by CUORE, presumed to be a muon. Colored indicates that the crystal is included in the event; red indicates saturation, blue indicates non-saturation. Figures both from \cite{yocum2022muon}}
            \label{fig:YocumTrackReconstruction}
        \end{figure}
        
        At mid-multiplicities (roughly 3-10 crystals per event), events are dominated by gammas --- either high energy gammas scattered across the detector, or events that produce more than one gamma with well-defined angular correlations. One such example is the excited state decay of \Te to \XeES which releases on average 2.12 \ga per event. Recent analyses have coded events not only based on energy but on geometrical associations between activated crystals\cite{Adams:2021}. A CNN trained on Monte Carlo data could identify such associations. Neural network classification algorithms already in use by other experiments \cite{Abbasi:2021,Aurisano:2016,Acciarri:2017} may aid in these events, especially when applied to the uniquely compact, segmented, calorimetric detector like \cupid \& \cupidt, as well as those with larger patterns across the detector like tri-nucleon decay.

\subsection{Quantum Computing} \label{sec:BroaderImpacts_QuantumComputing}
The next generation of quantum computers and superconducting qubit instruments must be operated in cryogenic infrastructure \cite{the_national_academies_of_sciences_engineering_and_medicine_quantum_2018}, read out thousands of channels, and shield the instrumentation from ionizing radiation. In fact, recent research published in Nature suggests that ionizing radiation from cosmic rays and natural radioactivity will be the next limiting factor in qubit coherence \cite{vepsalainen_impact_2020}. Thus, the next leap in QIS will require infrastructure that can successfully operate at sub-Kelvin temperatures at ultra-high vacuum with stringent radiopurity requirements — systems that can be developed by drawing from existing searches for rare-events in nuclear physics such as the search for \ndbd\ with CUPID-1T.
In the last years, several groups began to investigate the effects of environmental radioactivity on one of the most advanced technologies for the deployment of coherent quantum processors: superconducting circuits. In 2018, the DEMETRA collaboration\cite{cardani_JLTP_2020} proposed the idea that the performance of superconducting circuits could be spoiled by the absorption of energy in the substrate on which they are deposited. 
Today, there are increasing evidences that environmental radioactivity will (i) limit the coherence of next-generation ``transmon" qubits at the level of milliseconds~\cite{Vepsalainen2020} and (ii) induce correlated errors in matrix of qubits, undermining the most popular protocols for quantum error correction\cite{wilen_2021,McEwen_2021}.
On the other hand, there are encouraging studies proving that the operation of superconducting circuits in radio-pure environment allows to improve the quality factor of resonators\cite{Cardani_NatComm2021} and the locking of a gradiometric fluxonium at its sweet spot\cite{Gusenkova:2022}. 
The high specialization of the CUPID collaboration in the deployment of low-background cryogenic facilities in deep underground laboratories will allow to create the ideal environment for the operation of next-generation quantum processors.

\section{Summary and Outlook}
    The requirements and goals of a \cupidt experiment will ultimately necessitate a robust exchange of ideas and applications across scientific fields, in basic principles, instrumentation, data management techniques, and experimental infrastructure.  
        The technology and innovation required to produce a sensitive, ton-scale bolometric detector for \ndbd is under active development, and achievable within the next decade.  The overlapping, synergistic requirements of nuclear \dbd searches and searches for rare events and new particles in high energy physics highlight the importance of capitalizing on the expertise and goals of both communities to assemble and operate a versatile detector for rare event searches.

        \cupidt{} is an exciting ton-scale concept to search for neutrinoless double-beta decay and beyond. Current R\&D efforts to push development of the background-reduction techniques and readout capabilities necessary to realize the detector are underway. With current projections, \cupidt{} could begin construction as early as the late 2020's, and commissioning in the early 2030's. The development of systems for the stable multiplexing and readout of large arrays of macrobolometers, and continued innovation in low-background techniques and facilities, are key components for truly ton-scale searches for \ndbd{}, and important areas of overlap between Snowmass frontiers.

        \section{Acknowledgements}
        The CUPID Collaboration thanks the directors and staff of the Laboratori Nazionali del Gran Sasso and the technical staff of our laboratories. This work was supported by the Istituto Nazionale di Fisica Nucleare (INFN); by the European Research Council (ERC) under the European Union Horizon 2020 program (H2020/2014-2020) with the ERC Advanced Grant no. 742345 (ERC-2016-ADG, project CROSS) and the Marie Sklodowska-Curie Grant Agreement No. 754496; by the Italian Ministry of University and Research (MIUR) through the grant Progetti di ricerca di Rilevante Interesse Nazionale (PRIN 2017, grant no. 2017FJZMCJ); by the US National Science Foundation under Grant Nos. NSF-PHY-1401832, NSF-PHY-1614611, and NSF-PHY-1913374. This material is also based upon work supported by the US Department of Energy (DOE) Office of Science under Contract Nos. DE-AC02-05CH11231 and DE-AC02-06CH11357; and by the DOE Office of Science, Office of Nuclear Physics under Contract Nos. DE-FG02-08ER41551, DE-SC0011091, DE-SC0012654, DE-SC0019316, DE-SC0019368, and DE-SC0020423. This work was also supported by the Russian Science Foundation under grant No. 18-12-00003 and the National Research Foundation of Ukraine under Grant No. 2020.02/0011. This research used resources of the National Energy Research Scientific Computing Center (NERSC). This work makes use of both the DIANA data analysis and APOLLO data acquisition software packages, which were developed by the CUORICINO, CUORE, LUCIFER and CUPID-0 Collaborations.

\clearpage
~


%
\bibliographystyle{hunsrt}
\nocite{cupido2019}
\bibliography{document.bib}
\newpage

\vspace{3.5in}

\noindent {\large \bf Authors:} 
The CUPID Collaboration \\[+1em]

\author{A.~Armatol}
\affiliation{IRFU, CEA, Universit\'e Paris-Saclay, Saclay, France}

\author{C.~Augier}
\affiliation{Institut de Physique des 2 Infinis, Lyon, France}

\author{F.~T.~Avignone~III}
\affiliation{University of South Carolina, Columbia, SC, USA}

\author{O.~Azzolini}
\affiliation{INFN Laboratori Nazionali di Legnaro, Legnaro, Italy}

\author{M.~Balata}
\affiliation{INFN Laboratori Nazionali del Gran Sasso, Assergi (AQ), Italy}

\author{K.~Ballen}
\affiliation{INFN Laboratori Nazionali di Legnaro, Legnaro, Italy}

\author{A.~S.~Barabash}
\affiliation{National Research Centre Kurchatov Institute, Institute for Theoretical and Experimental Physics, Moscow, Russia}

\author{G.~Bari}
\affiliation{INFN Sezione di Bologna, Bologna, Italy}

\author{A.~Barresi}
\affiliation{INFN Sezione di Milano - Bicocca, Milano, Italy}
\affiliation{University of Milano - Bicocca, Milano, Italy}

\author{D.~Baudin}
\affiliation{IRFU, CEA, Universit\'e Paris-Saclay, Saclay, France}

% ...

\end{document}